\documentclass[a4paper, fleqn, usenatbib]{mnras}
\usepackage{newtxtext,newtxmath}
\usepackage[T1]{fontenc}
\usepackage{ae,aecompl}
\usepackage{graphicx}	% Including figure files
\usepackage{amsmath}	% Advanced maths commands
\usepackage{amssymb}	% Extra maths symbols
\usepackage{times}

\def \aj {AJ}
\def \mnras {MNRAS}
\def \pasp {PASP}
\def \apj {ApJ}
\def \apjs {ApJS}
\def \apjl {ApJL}
\def \aap {A\&A}

\def \araa {ARAA}

\def \aaps {A\&A Suppl.}
\def \pasa {PASA}

\def\lesssim{\mathrel{\hbox{\rlap{\hbox{\lower4pt\hbox{$\sim$}}}\hbox{$<$}}}}
\def\gtrsim{\mathrel{\hbox{\rlap{\hbox{\lower4pt\hbox{$\sim$}}}\hbox{$>$}}}}

\long\def\symbolfootnote[#1]#2{\begingroup%
\def\thefootnote{\fnsymbol{footnote}}\footnote[#1]{#2}\endgroup}

\title[Type IIP SN stellar populations]{The resolved stellar populations around 12 Type IIP supernovae}

\author[Maund]{Justyn R. ~Maund\thanks{email: j.maund@sheffield.ac.uk}\thanks{Royal Society Research Fellow}\\
$$ Department of Physics and Astronomy, University of Sheffield, Hicks Building, Hounsfield Road, Sheffield S3 7RH, U.K.\\
}

\date{Accepted XXX. Received YYY; in original form ZZZ}

\pubyear{2016}

\begin{document}
\label{firstpage}
\pagerange{\pageref{firstpage}--\pageref{lastpage}}
\maketitle
%%%%%%%%%%%%%%%%%%%%%%%%%%%%%%%%
%ABSTRACT
%ABSTRACT
%ABSTRACT
%%%%%%%%%%%%%%%%%%%%%%%%%%%%%%%%

\begin{abstract}

Core-collapse supernovae are found in regions associated with recent massive star formation.  The stellar population observed around the location of a SN can be used as a probe of the origins of the progenitor star.   We apply a Bayesian mixture model to fit isochrones to the massive star population around twelve Type IIP SNe, for which constraints on the progenitors are also available from fortuitous pre-explosion images.  Using the high-resolution Hubble Space Telescope Advanced Camera for Surveys and Wide Field Camera 3,  we study the massive star population found within 100pc of each our target SNe.   For most of the SNe in our sample, we find that there are multiple age components in the surrounding stellar populations.  In the cases of SNe~2003gd and 2005cs, we find that the progenitor does not come from the youngest stellar population component and, in fact, these relatively low mass progenitors ($\sim 8M_{\odot}$) are found in close proximity to stars as massive as $15$ and $50-60M_{\odot}$, respectively.  Overall, the field extinction (Galactic and host) derived for these populations is $\sim 0.3\,\mathrm{mags}$ higher than the extinction that was generally applied in previously reported progenitor analyses.  We also find evidence, in particular for SN~2004dj, for significant levels of differential extinction.
Our analysis for SN~2008bk suggests a significantly lower extinction for the population than the progenitor, but the lifetime of the population and mass determined from pre-explosion images agree.
Overall,  assuming that the appropriate age component can be suitably identified from the multiple stellar population components present, we find that our Bayesian approach to studying resolved stellar populations can match progenitor masses determined from direct imaging to within $\pm 3M_{\odot}$.

\end{abstract}
%%%%%%%%%%%%%%%%%%%%%%%%%%%%%%%%
%KEYWORDS
%KEYWORDS
%KEYWORDS
%%%%%%%%%%%%%%%%%%%%%%%%%%%%%%%%

%%%%%%%%%%%%%%%%%%%%%%%%%%%%%%%%
%INTRODUCTION
%INTRODUCTION
%INTRODUCTION
%%%%%%%%%%%%%%%%%%%%%%%%%%%%%%%%
\section{Introduction}
\label{sec:intro}
The search for the progenitors of core-collapse supernovae (CCSNe) has been
particularly successful in identifying the red supergiant (RSG) progenitors responsible for the hydrogen rich Type IIP SNe.    The identification of the progenitors of these events is limited to those SNe with fortuitous pre-explosion images.  Such direct detections of the progenitors have yielded initial mass estimates for the exploding stars, that directly link the processes of stellar evolution with the properties of the subsequent SNe.  \citet{2008arXiv0809.0403S} reported that the ensemble of progenitor mass estimates indicated that the most massive RSGs (arising from stars with masses $>16M_{\odot}$) were not exploding as Type IIP SNe.  The quality of the pre-explosion observations, however, requires some caution in the interpretation of the initial mass estimates.  Limited pre-explosion detections, in only one or two filters, are insufficient to determine the properties of the progenitors without requiring significant assumptions such as: the temperature, reddening (Galactic, host and local to the progenitor), the choice of reddening law and metallicity \citep{2014MNRAS.438..938M}.  In addition, the poor quality of the pre-explosion observation often do not yield a detection of the progenitor, and so only limits on the initial mass can be derived.  While non-detections make up a significant fraction of the progenitor constraints for Type IIP SNe \citep{2008arXiv0809.0403S}, the situation is dramatically worse for the Type Ibc SNe \citep{2013MNRAS.436..774E} for which there has only been one confirmed detection of a progenitor \citep{2013ApJ...775L...7C,2016MNRAS.461L.117E,2016ApJ...825L..22F}.

Given the limitations concerning the availability and quality of appropriate pre-explosion data, and restrictions of the subsequent analysis and the inherent large uncertainties (both systematic and statistical), alternative proxies for the progenitor have been sought through analysis of the host environment.  Under the assumption that the short-lived, massive stars that die as CCSNe will explode close to the regions in which they are born, and that the majority of such stars are born in proximity to other massive stars \citep{2016ApJ...817..113L},  the apparent age of the environment hosting a SN may be expected to correspond to the lifetime of the progenitor and, hence, a measure of the initial mass.  Many different approaches have been used to characterise these host environments including association with {\sc H ii} regions, the measured equivalent with of $H\alpha$ at the SN position (or, more usually, the nearest strong $H\alpha$ emission) \citep[e.g.][]{2013AJ....146...30K, 2013AJ....146...31K} and the use of pixel statistics methods \citep[for a review see][]{2015PASA...32...19A}, which assess the relative brightness of the host galaxy at the SN position at different wavelengths that serve as indicators for recent star formation.  These techniques, however, provide indirect measures of the properties of the host stellar population, and may gloss over intricate physical processes concerning the relationship between massive stars and {\sc H ii} regions, the evolution of massive stars, their spatial distribution and the age composition of the stellar population \citep{2013MNRAS.428.1927C}.

In some instances, such as SNe 2004dj and 2009kr \citep{2004ApJ...615L.113M, 2005ApJ...626L..89W,2009ApJ...695..619V,2015MNRAS.447.3207M}, host clusters have been identified at the pre-explosion position of the SN, for which ages have been directly derived through the comparison of the observed spectral energy distributions (SEDs) with cluster models such as {\sc starburst99} \citep{1999ApJS..123....3L}.  Such clusters, however, represent only one end of the spectrum of stellar densities and, in general, the massive star environment associated with most CCSNe are consistent with looser associations \citep{2011MNRAS.410L...6G}.  

Previously, \citet{1999AJ....118.2331V} and \citet{2005astro.ph..1323M} had presented photometry of the resolved environments surrounding nearby SNe, as imaged by the Hubble Space Telescope (HST), providing qualitative interpretations of the corresponding age of the resolved stellar population.   \citet{2009ApJ...703..300G} compared the observed photometry of the stellar population around the  2008 optical transient in NGC~300, against synthetic colour-magnitude diagrams (CMDs) for a range of star formation histories to determine the mass of the progenitor; and using the same technique \citet{2011ApJ...742L...4M} was able to determine a mass for the progenitor of SN~2011dh that was comparable to the mass derived for the progenitor itself in pre-explosion observations \citep{2011ApJ...739L..37M,2011ApJ...741L..28V}.  As this technique is not reliant on pre-explosion observations, further studies have used SN remnants to identify stellar populations associated with SNe and provide initial mass estimates for the progenitors of these unobserved events \citep{2009ApJ...700..727B,2012ApJ...761...26J,2014ApJ...795..170J}.
\citet{2016MNRAS.456.3175M} recently used an alternative strategy, using a Bayesian isochrone fitting technique that simultaneously used all seven-colour photometry of the host environment, to determine the properties of the stellar population around the position of the Type Ic SN~2007gr in NGC~1058. Despite the absence of the progenitor in pre-explosion $WFPC2$ observations, from their analysis of the resolved stellar population \citeauthor{2016MNRAS.456.3175M} concluded that star that exploded was not a low mass star stripped by a binary companion, but rather a high mass star $M_{init}\sim 40M_{\odot}$ that had probably undergone a Wolf-Rayet phase.

\citet{2014ApJ...791..105W} presented a study of the resolved stellar populations associated with 17 historical observed SNe.  Despite relatively large uncertainties, they concluded that the ages of stellar populations around the sites of those 5 SNe with identified progenitors yielded mass constraints consistent with those derived from pre-explosion images.  

Here we present a study of the stellar populations associated with 12 recent Type IIP SNe, as observed by HST, for which progenitor candidates or host clusters have been identified in pre-explosion observations.  A list of the SNe considered here is presented in Table \ref{tab:obs:samp}.  The aim of this study is to ascertain the usefulness of the Bayesian isochrone fitting technique to determine ages and extinctions for the surrounding stellar populations of the target SNe, for the direct comparison with the progenitor properties directly inferred from pre-explosion observations.

%%%%%%%%%%%%%%%%%%%%%%%%%%%%
%TABLE OF SAMPLE SUPERNOVA
%%%%%%%%%%%%%%%%%%%%%%%%%%%%

\begin{table*}
\caption{\label{tab:obs:samp} Sample of 12 Type IIP Supernovae with HST observations.}
\begin{tabular}{ccccc}
\hline
	Supernova  	& Host Galaxy  & $m - M$ 			& $A_{V}$  		& $12 + \log \left(O/H\right)$\\
				&			& (mags)				&   (mags)$^{1}$ 	& 					\\ 
\hline
	2003gd$^{2,3}$	&	M74	   	& 	29.84 (0.42)$^{12}$	&	0.192 		& 8.4$^{22}$\\
	2004A$^{3}$	&	NGC 6207&	31.38 (0.07)$^{13}$	&	0.042		& 8.3$^{22}$\\
	2004dj$^{4}$	&	NGC 2403&	27.51 (0.07)$^{14}$	&	0.110		& 8.4$^{22}$\\
	2004et$^{5}$	&	NGC 6946&	28.46 (0.06)$^{15}$	&	0.938		&8.3$^{22}$\\
	2005cs$^{3}$	&	M51	   	& 	29.26 (0.37)$^{16}$	&	0.095		&8.7$^{22}$\\
	2006ov$^{5,6}$	&	NGC 4303&	30.51 (0.09)$^{17}$	&	0.061		&8.9$^{22}$\\
	2008bk$^{7}$	&	NGC 7793&	27.74 (0.06)$^{18}$	&	0.053		&8.4$^{7}$\\
	2009kr$^{8}$	&	NGC 1832&	32.09 (0.15)$^{19}$	&	0.200		&8.61$^{8}$\\
	2009md$^{8}$	&	NGC 3389&	31.63 (0.55)$^{19}$	&	0.074		&8.6$^{8}$\\
	2012aw$^{9}$	&	NGC 3351&	30.00 (0.04)$^{20}$	&	0.076		&8.8$^{9}$\\
	2012ec$^{10}$	&	NGC 1084&	31.21 (0.10)$^{21}$	&	0.073		&8.93$^{10}$\\
	2013ej$^{11}$	&	M74	   	& 	29.84 (0.42)$^{12}$	&	0.192		&8.6$^{11}$\\
	\hline
\end{tabular}\\
$^{1}$ Foreground extinction from the dust maps of \citet{2011ApJ...737..103S} as quoted by the NASA/IPAC Extragalactic Database;
	$^{2}$ \citet{2009Sci...324..486M};
	$^{3}$ \citet{2014MNRAS.438..938M};
	$^{4}$ \citet{2009ApJ...695..619V};
	$^{5}$ \citet{2011MNRAS.410.2767C};
	$^{6}$ \citet{2007ApJ...661.1013L};
	$^{7}$ \citet{2014MNRAS.438.1577M};
	$^{8}$ \citet{2015MNRAS.447.3207M};
	$^{9}$ \citet{2012arXiv1204.1523F};
	$^{10}$ \citet{2013arXiv1302.0170M};
	$^{11}$ \citet{2014MNRAS.439L..56F};
	$^{12}$ \citet{2005MNRAS.359..906H};
	$^{13}$ Weighted average of distances derived for SN 2004A \citep{2015ApJ...799..215P,2014AJ....148..107R} and Tully-Fisher distances to NGC~6207 \citep{2014MNRAS.444..527S,2009AJ....138..323T, 2013AJ....146...86T,2009ApJS..182..474S};
	$^{14}$ \citet{2011ApJS..195...18R};
	$^{15}$ Weighted average of Expanding Photosphere Method distances determined for SN~2004et \citep{2014ApJ...782...98B,2012MNRAS.419.2783T};
	$^{16}$ \citet{2006MNRAS.372.1735T};
	$^{17}$ Weighted average of distances determined to SN 2008in \citep{2011ApJ...736...76R, 2014ApJ...782...98B}
	and Tully-Fisher distance \citep{1997AA...323...14S};
	$^{18}$ \citet{2010AJ....140.1475P,2009AJ....138..332J};
	$^{19}$ see \citet{2015MNRAS.447.3207M} and references therein;
	$^{20}$ Weighted average of distances determined from SN~2012aw \citep{2014ApJ...782...98B, 2015MNRAS.448.2312B, 2015AA...580L..15P, 2015ApJ...799..215P,2014AJ....148..107R};
	$^{21}$ Weighted average of distances determined from SN~2012ec \citep{2015MNRAS.448.2312B}; $^{22}$ \citep{2008arXiv0809.0403S}.
	
\end{table*}
%%%%%%%%%%%%%%%%%%%%%%%%%%%%%
%OBSERVATIONS
%OBSERVATIONS
%OBSERVATIONS
%%%%%%%%%%%%%%%%%%%%%%%%%%%%%

\section{Observations and Data Reduction}
\label{sec:obs}
The data for the 12 target SNe was retrieved from the Barbara A. Mikulski Archive for Space Telescopes\footnote{https://archive.stsci.edu/hst/}; having been processed via the on-the-fly recalibration pipeline with the latest calibrations.  To ensure uniformity across the sample we only selected those observations acquired with the Wide Field Camera 3 Ultraviolet-Visible (WFC3 UVIS) channel and the Advanced Camera for Surveys (ACS) High Resolution Channel (HRC) and Wide Field Channel (WFC).  A log of these observations is presented in Table \ref{tab:obs:obs}.  To determine the location of the SN on these observations, we used the positions reported in previous analyses of the progenitor candidates identified in pre-explosion observations (see references in Table \ref{tab:obs:samp}).

Photometry of the observations was conducted using the {\sc dolphot} package\footnote{http://americano.dolphinsim.com/dolphot/} \citep{dolphhstphot},  with the instrument specific modules, following the prescription of \citet{2016MNRAS.456.3175M}.  We consider all sources within a circular area around each SN position with a radius $\mathrm{100\,pc}$.  \citet{2009ApJ...703..300G} suggested ${\mathrm 50\,\mathrm{pc}}$ was sufficient to probe the stellar population they claim to be most closely associated with the progenitor.  They based their decision on the available observations of the nearby galaxy NGC~300, however our sample extends to distances a factor $\sim 15$ larger.  We also note that their sample radius was chosen to limit background contamination, however \citet{2016MNRAS.456.3175M} found an approximately coeval population of SN 2007gr extended to $\sim 150\,\mathrm{pc}$.   The final photometric catalogues were limited to those stars, in the vicinity of the SN, that were detected in at least one filter with a $S/N \geq 5$ and were classified by {\sc dolphot} as being of type 1 (nominally stellar or point-like).  
For stars that were not detected in all available filters, detection limits were determined using artificial star tests.  An artificial star was considered to have been successfully recovered if it was found within $1\,\mathrm{pix}$ of the inserted position, was detected at $S/N\geq 5$ and the difference between the input and recovered brightness was less than the photometric uncertainty.  The detection probability function was parameterised as a cumulative normal distribution function.  We also used artificial star tests, randomly positioned within the $\mathrm{100\,pc}$ circular region to assess additional systematic photometric uncertainties due to crowding.    

%The isochrone fitting technique we use is insensitive to whether a given star is in the background or not (automatically assigning such stars to the appropriate age component).

\begin{table*}
	\caption{HST observations of the environments around the sample of 12 Type IIP SNe \label{tab:obs:obs}}
\begin{tabular}{ccccccccc}
\hline
	SN	& {Dataset}	& {Date}& {Instrument}	&{Detector/Aperture}	& {Filter}&{Exposure}	& {Pixel}			& {Program}\\
		&		&	(UT)	&					&				&		&Time (s)		& Size ($\arcsec$)	& \\
\hline
	2003gd&JB4T02010&2010 Nov 14&ACS   &WFC1-1K&F555W&1364&0.05&11675$^{1}$\\
	&JB4T02020&2010 Nov 14&ACS   &WFC1-1K&F814W&1398&0.05&11675\\
	&JB4T02030&2010 Nov 14&ACS   &WFC1-1K&F435W&1600&0.05&11675\\
	\\
	2004A&JB4T03010&2010 Sep 09&ACS   &WFC1-1K&F555W&1400&0.05&11675$^{1}$\\
	&JB4T03020&2010 Sep 09&ACS   &WFC1-1K&F814W&1434&0.05&11675\\
	&JB4T03030&2010 Sep 09&ACS   &WFC1-1K&F435W&1636&0.05&11675\\
	\\
	2004dj&J9F005010&2005 Aug 28&ACS   &HRC&F606W&1648&0.025&10607$^{2}$\\
	&J9F005030&2005 Aug 28&ACS   &HRC&F814W&1600&0.025&10607\\
	\\
	2004et&IC8S01010&2014 Feb 20&WFC3  &UVIS1&F438W&2768&0.04&13392$^{2}$\\
	&IC8S01020&2014 Feb 20&WFC3  &UVIS1&F606W&1744&0.04&13392\\
	&IC8S01030&2014 Feb 20&WFC3  &UVIS1&F814W&1744&0.04&13392\\
	\\
	2005cs&JB4T04010&2010 Jul 30&ACS   &WFC1-1K&F555W&1460&0.05&11675$^{1}$\\
	&JB4T04020&2010 Jul 30&ACS   &WFC1-1K&F814W&1494&0.05&11675\\
	&JB4T04030&2010 Jul 30&ACS   &WFC1-1K&F435W&1696&0.05&11675\\
	\\
	2006ov&JBS801010&2012 May 24&ACS   &WFC1-CTE&F435W&1090&0.05&12574$^{3}$\\
	&JBS801020&2012 May 24&ACS   &WFC1-CTE&F814W&1090&0.05&12574\\
	\\
	2008bk&IBHH02010$^{\dagger}$&2011 Apr 29&WFC3  &UVIS2-C1K1C-SUB&F814W&915&0.04&12262$^{1}$\\
	&IBHH02020$^{\dagger}$&2011 Apr 29&WFC3  &IRSUB512&F160W&828.577&0.13&12262\\
	&IBHH02030$^{\dagger}$&2011 Apr 30&WFC3  &IRSUB512&F125W&461.837&0.13&12262\\
	&IBKU01030&2011 May 07&WFC3  &UVIS1&F336W&1506&0.04&12285$^{4}$\\
	&IBKU01040&2011 May 07&WFC3  &UVIS1&F555W&1506&0.04&12285\\
	&IBKU01050&2011 May 07&WFC3  &UVIS1&F814W&1512&0.04&12285\\
	\\
	2009kr&JBQ701010&2012 Oct 26&ACS   &WFC1-1K&F814W&1536&0.05&12559$^{1}$\\
	&JBQ701020&2012 Oct 26&ACS   &WFC1-1K&F435W&1080&0.05&12559\\
	&JBQ701030&2012 Oct 26&ACS   &WFC1-1K&F555W&1700&0.05&12559\\
	\\
	2009md&JBQ702010&2012 Nov 08&ACS   &WFC1-1K&F814W&1536&0.05&12559$^{1}$\\
	&JBQ702020&2012 Nov 08&ACS   &WFC1-1K&F435W&1072&0.05&12559\\
	&JBQ702030&2012 Nov 08&ACS   &WFC1-1K&F555W&1700&0.05&12559\\
	\\
	2012aw&JCJ601010&2015 Mar 22&ACS   &WFC1-1K&F555W&3268&0.05&13825$^{1}$\\
	&JCJ601020&2015 Mar 22&ACS   &WFC1-1K&F814W&1536&0.05&13825\\
	\\
	2012ec&JCUC01010&2016 Jul 30&ACS   &WFC1-1K&F475W&1560&0.05&14226$^{1}$\\
	&JCUC01020&2016 Jul 30&ACS   &WFC1-1K&F606W&940&0.05&14226\\
	&JCUC01030&2016 Jul 30&ACS   &WFC1-1K&F814W&1728&0.05&14226\\
	\\
	2013ej&J8OL04010&2003 Nov 20&ACS   &WFC1&F814W&1560&0.05&9796$^{5}$\\
	&J8OL04020&2003 Nov 20&ACS   &WFC1&F555W&2200&0.05&9796\\
	&J8OL04030&2003 Nov 20&ACS   &WFC1&F435W&2360&0.05&9796\\
	&J8OL04050&2003 Nov 20&ACS   &WFC1&F435W&2360&0.05&9796\\
\hline
\end{tabular}\\
$^{\dagger}$ Data not included in full analysis, see Section \ref{sec:res:08bk};$^{1}$ PI: J.R. Maund; $^{2}$ B. Sugerman; $^{3}$ D. Leonard; $^{4}$ R. Soria; $^{5}$ J. Miller.
\end{table*}

%%%%%%%%%%%%%%%%%%%%%%%%%%%%%%%%
%RESULTS AND ANALYSIS
%RESULTS AND ANALYSIS
%RESULTS AND ANALYSIS
%%%%%%%%%%%%%%%%%%%%%%%%%%%%%%%%
\section{Results \& Analysis}
\label{sec:res}

\begin{figure*}
\includegraphics[width=18cm]{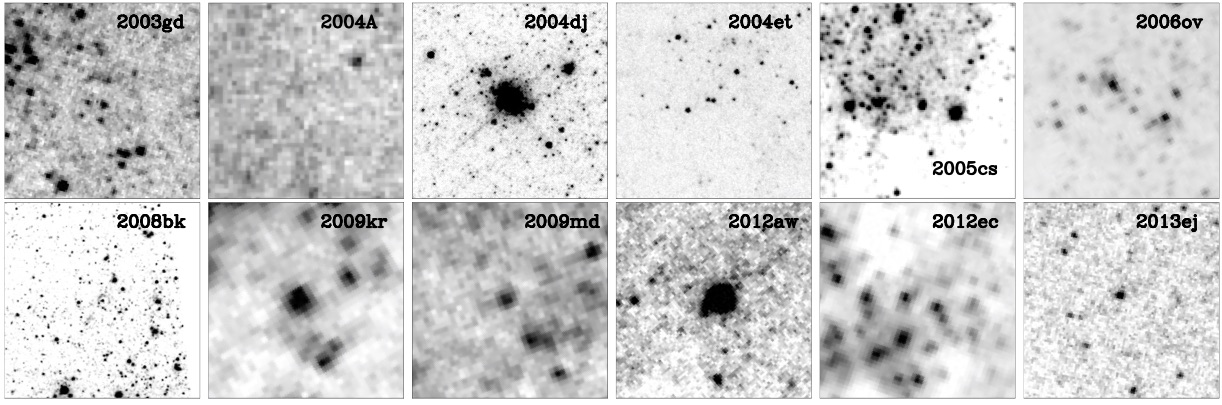}
\caption{Hubble Space Telescope observations of the sites of 12 Type IIP Supernovae (all images are $\sim V$-band observations acquired with the $F555W$ or $F606W$ filters, except for SN~2006ov for which we have used an $F814W$ image).  Each image stamp is centred on the location of the SN (and oriented with North up and East to the left) and has dimensions of $200\,\mathrm{pc}\,\times 200\,\mathrm{pc}$ at the distances of the respective host galaxies. \label{fig:res:img}}
\end{figure*}

\begin{figure}
\begin{center}
\includegraphics[width=6.5cm, angle=270]{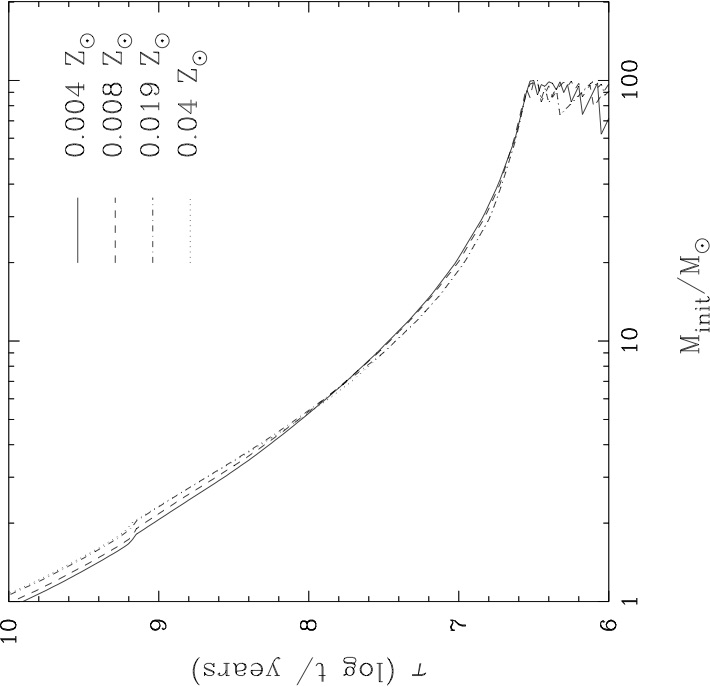}
\vspace{0.2cm}
\end{center}
\caption{Progenitor lifetime as a function of initial mass and metallicity, from the Padova stellar evolution models \citep{2002A&A...391..195G,2008A&A...482..883M}.}
\label{fig:res:massage}
\end{figure}

The sites of the Type IIP SNe considered in our sample, as imaged by  HST, are presented in Figure \ref{fig:res:img}.  We applied the technique of \citet{2016MNRAS.456.3175M} to analyse the stellar populations around the SNe with respect to Padova isochrones \citep{2002A&A...391..195G,2008A&A...482..883M}\footnote{http://pleiadi.pd.astro.it/}, selected for the appropriate metallicity (see Table \ref{tab:obs:samp}) assuming the metallicity scales with oxygen abundence (for which we adopt the solar oxygen abundance of $12 + \log(O/H) = 8.69\pm0.05$; \citealt*{2009ARA&A..47..481A}).  The ages determined from our analysis were assumed to correspond to the lifetimes of the progenitors, from which initial masses were determined (see Figure \ref{fig:res:massage}).

The isochrone fitting technique presented by \citet{2016MNRAS.456.3175M}, following \citet{2005A&A...436..127J}, uses a Bayesian mixture model to fit multiple age components $\tau = \log (t /\mathrm{years})$, assigning each component a weight $w$ representing the fraction of the stellar population belonging to that component.   For the prior probability for the initial mass of each star we follow  \citeauthor{2016MNRAS.456.3175M} and adopt a \citet{1955ApJ...121..161S} Initial Mass Function (IMF). The technique uses the Nested Sampling algorithm \citep{2004AIPC..735..395S,2008MNRAS.384..449F,2013arXiv1306.2144F}, such that we adopt the Bayes factor (the ratio of Bayesian evidence/marginal likelihoods $Z$) to select the appropriate model and number of age components $N_{m}$ and, simultaneously, derive the posterior probability distributions for the parameters of the fit.  The Bayes factor $K$ for comparing models with $n$ and $n + 1$ components is given, therefore, by:
\begin{equation}
K(n+1 , n) = \frac{B_{n+1}}{B_{n}} = \exp\left( \ln \left(Z_{n+1}/(n+1)!\right) - \ln \left(Z_{n}/n!\right)\right)
\end{equation}
where the factorial terms correct the marginal likelihoods for mirror modes, due to the interchangeability of the each of the age compoints.  In interpreting the final values of $K$ we follow the significance scale of \citet{Jeffreys61}, although we note that \citet{2008ConPh..49...71T} suggest that this scale may be too conservative.  Further to \citeauthor{2016MNRAS.456.3175M},  we note that once the optimal value of $N_{m}$ is achieved the addition of more age components will not result in any additional improvement in the value of $Z$ (and values of $K$ will decrease).  In order to provide a consistent analysis across the sample, we have chosen to restrict the maximum number of mixture components to 3.

Unlike \citeauthor{2016MNRAS.456.3175M}, in this study we also include the effects of differential extinction $\mathrm{d}A_{V}$, such that the assumed extinction towards individual members of the stellar populations is drawn from a distribution $\sim N\left(A_{V}, (dA_{V})^{2}\right)$.   We adopt a logarithmic prior over the range $0 \leq \log_{10}\left(\mathrm{d}A_{V}/0.02\right) \leq 1$, penalising large values of differential extinction.  For the extinction, we assume a standard Galactic \citet{ccm89} $R_{V} = 3.1$ reddening law.  We also fix the width of each age component to $\sigma_{\tau} = 0.050$ to limit the effect of degeneracies between age and extinction in fitting CMDs with observations in only two filters.  If a stellar population component requires a broader age, this is naturally handled during the fitting process through additional, but adjacent, age components. 

A summary of the results of the application of our Bayesian stellar populations analysis to our sample of 12 Type IIP SNe is presented in Table \ref{tab:res}; the results for individual objects are discussed in the following sections.


\begin{table*}
\caption{Results of the Bayesian stellar populations analysis for the sample of 12 Type IIP Supernovae\label{tab:res}}
\bgroup
\def\arraystretch{1.5}% 
\begin{tabular}{lccccccccccc}
\hline
SN		& $N_{\star}$	& $\ln K(2, 1)$	& $\ln K(3, 2)$	& $\tau_{1}$	& $w_{1}$		&$\tau_{2}$	& $w_{2}$		&$\tau_{3}$	& $w_{3}$	& $A_{V}$		& $\mathrm{d}A_{V}$	\\
\hline
2003gd &	161     &$107.43  (0.21)$ & $-0.13 (0.24)$ &	$7.14_{-0.03}^{+0.06}$ &	0.32 &	$8.08_{-0.02}^{+0.02}$ &	0.68 &	\ldots                  &	\ldots &	$0.46_{-0.03}^{+0.04}$ &	0.04 \\	
2004A  &	59      &$11.49   (0.23)$ & $-0.58 (0.23)$ &	$6.31_{-0.22}^{+0.37}$ &	0.08 &	$7.74_{-0.04}^{+0.03}$ &	0.92 &	\ldots                  &	\ldots &	$0.64_{-0.14}^{+0.09}$ &	0.19 \\
2004dj &	100     &$24.62   (0.26)$ & $-2.32 (0.27)$ &	$7.22_{-0.04}^{+0.05}$ &	0.58 &	$8.11_{-0.04}^{+0.02}$ &	0.42 &	\ldots                  &	\ldots &	$0.94_{-0.14}^{+0.11}$ &	0.35 \\	
2004et &	240     &$1080.03 (0.26)$ & $98.66 (0.28)$ &	$7.04_{-0.04}^{+0.04}$ &	0.49 &	$8.32_{-0.01}^{+0.02}$ &	0.25 &	$9.41_{-0.02}^{+0.01}$      &	0.27   &	$1.20_{-0.10}^{+0.10}$ &	0.06\\
2005cs &	385     &$884.00  (0.31)$ & $201.90(0.35)$ &	$6.35_{-0.02}^{+0.02}$ &	0.42 &	$7.21_{-0.02}^{+0.02}$ &	0.19 &	$7.63_{-0.01}^{+0.01}$      &	0.38   &	$0.54_{-0.01}^{+0.01}$ &	0.06\\	
2006ov &	48      &$206.00  (0.25)$ & $11.80 (0.28)$ &	$6.70_{-0.18}^{+0.04}$ &	0.48 &	$7.13_{-0.15}^{+0.08}$ &	0.10 &	$7.50_{-0.02}^{+0.03}$      &	0.42   &	$0.25_{-0.06}^{+0.08}$ &	0.06\\
2008bk &	316     &$1270.16 (0.32)$ & $161.24(0.34)$ &	$7.41_{-0.02}^{+0.04}$ &	0.32 &	$8.36_{-0.01}^{+0.01}$ &	0.49 &	$9.18_{-0.02}^{+0.02}$      &	0.20    &	$0.28_{-0.02}^{+0.02}$ &	0.10\\	
2009kr &	66      &$147.51  (0.29)$ & $5.55  (0.32)$ &	$6.77_{-0.02}^{+0.02}$ &	0.38 &	$7.15_{-0.02}^{+0.01}$ &	0.62 &	\ldots                  &   	\ldots &	$0.35_{-0.02}^{+0.03}$ &	0.06\\
2009md &	140     &$161.76  (0.27)$ & $33.85 (0.30)$ &	$6.71_{-0.03}^{+0.02}$ &	0.37 &	$7.23_{-0.02}^{+0.03}$ &	0.13 &	$7.49_{-0.02}^{+0.01}$      &	0.48   &	$0.55_{-0.04}^{+0.05}$ &	0.06\\	
2012aw &	83      &$75.71   (0.28)$ & $-3.06 (0.31)$ &	$6.19_{-0.06}^{+0.45}$ &	0.15 &	$7.93_{-0.01}^{+0.02}$ &	0.85 &	\ldots                  &	\ldots &	$1.37_{-0.13}^{+0.08}$ &	0.20\\	
2012ec &	394     &$83.75   (0.31)$ & $33.07 (0.33)$ &	$6.55_{-0.01}^{+0.01}$ &	0.44 &	$6.84_{-0.01}^{+0.02}$ &	0.18 &	$7.08_{-0.01}^{+0.01}$      &	0.36   &	$0.72_{-0.02}^{+0.05}$ &	0.20\\
2013ej &	185     &$127.41  (0.21)$ & $-1.19 (0.24)$ &	$7.16_{-0.03}^{+0.04}$ &	0.59 &	$8.03_{-0.02}^{+0.03}$ &	0.41 &	\ldots                  &	\ldots &	$0.45_{-0.04}^{+0.04}$ &	0.06\\
\hline
\end{tabular}
\egroup
\end{table*}
%%%%%%%%%%%%%%%%%%%%%%%%%%%%%%%%%%
%SUBSECTIONS
%SUBSECTIONS
%%%%%%%%%%%%%%%%%%%%%%%%%%%%%%%%%%
%2003gd
%2003gd
%2003gd
\subsection{SN 2003gd}
\begin{figure}
\begin{center}
\includegraphics[width=6.5cm, angle=270]{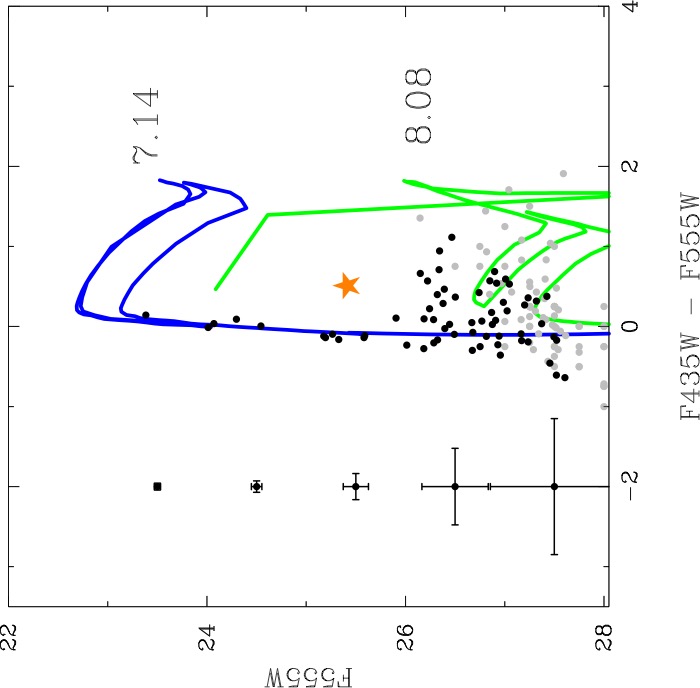}

\vspace{0.5cm}

\includegraphics[width=6.5cm,angle=270]{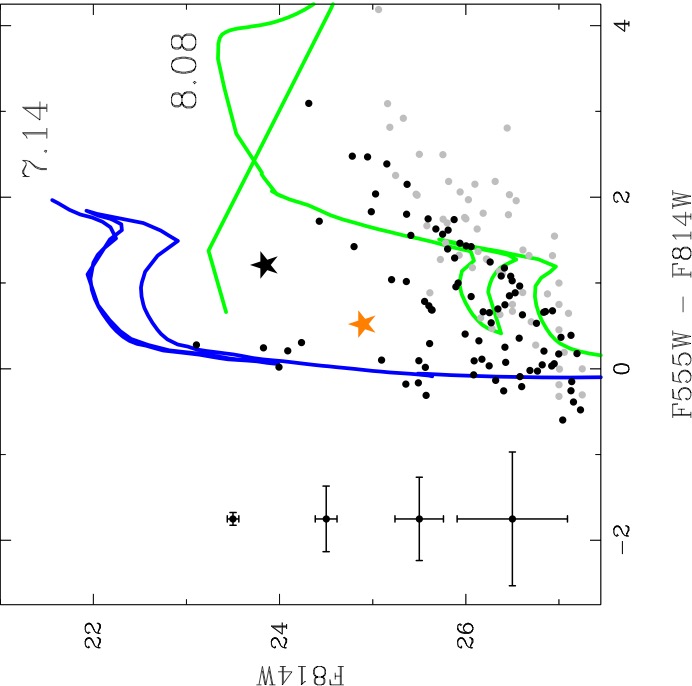}
\vspace{0.2cm}
\end{center}
\caption{CMDs for stars observed around the site of SN~2003gd in M74. The position of the progenitor detected in pre-explosion observations \citep{2014MNRAS.438..938M} is indicated by the black star ($\star$).  Black points indicate stars detected in both filters, while grey points indicate those stars detected in only one of the filters.  The column of points with error bars on the left-hand side of the plots shows the average uncertainties for stars at that brightness.  The orange star indicates the photometry of the late-time source recovered at the position of SN~2003gd in 2010 by \citet{2014MNRAS.438..938M}.}
\label{fig:cmd:03gd}
\end{figure}

CMDs for the population at the site of SN~2003gd are presented in Fig. \ref{fig:cmd:03gd}.   The SN is located in a region dominated by bright, blue stars, despite the progenitor itself being the sole obvious bright, red source in the pre-explosion observations \citep{2009Sci...324..486M}.  As \citet{2014MNRAS.438..938M} detected a faint blue source coincident with the position of the SN at late-times, we exclude it from our analysis.  There is very little scatter in the distribution of points in the ``blue plume", indicating a limited amount of differential extinction ($0.04\,\mathrm{mags}$) for this population.  The mean extinction and the amount of differential extinction towards the stellar population is consistent with the values derived by \citet{2014MNRAS.438..938M} in their analysis of the SEDs of the same stars in the vicinity of the SN (derived from the same 2010 HST observations).

The age of this population corresponds to the life-time of a star with $M_{init} \sim 14M_{\odot}$.  A fainter redder, plume indicates the presence of an older population, corresponding to a mass of $5M_{\odot}$ which is slightly lower than the mass estimated by \citet{2014MNRAS.438..938M} of $7.3\pm 1.9M_{\odot}$.  In considering the progenitor as a member of these populations, the location of the star in the CMD would appear to be almost intermediate between the two age components.  The progenitor photometry (the brightness in the pre-explosion WFPC2 $F606W$ and Gemini GMOS $i^{\prime}$ observations, without corrections for filter transformations) lies at the bright extreme of the older stellar population, yet significantly fainter than would be expected for a star that arose from the younger population.     
%2004A
%2004A
%2004A
\subsection{SN~2004A}
\label{sec:04A}
\begin{figure}
\begin{center}
\includegraphics[width=6.5cm, angle=270]{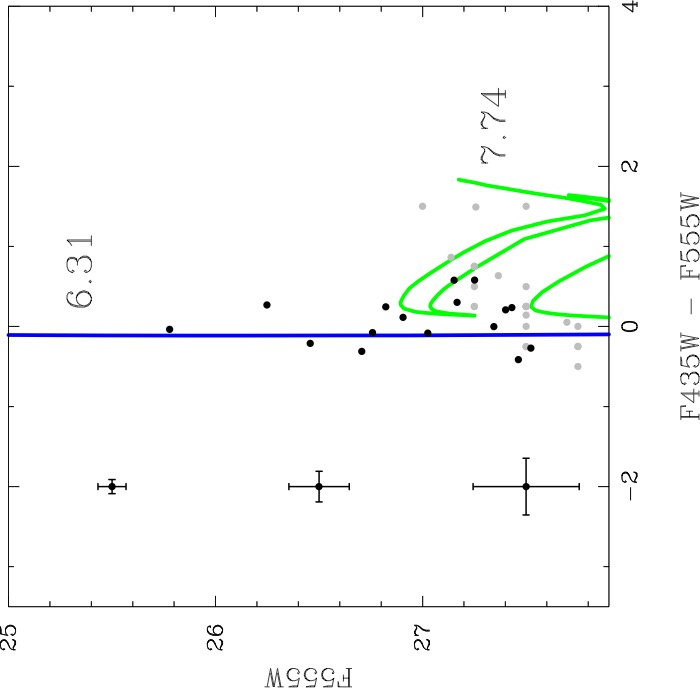}

\vspace{0.5cm}

\includegraphics[width=6.5cm,angle=270]{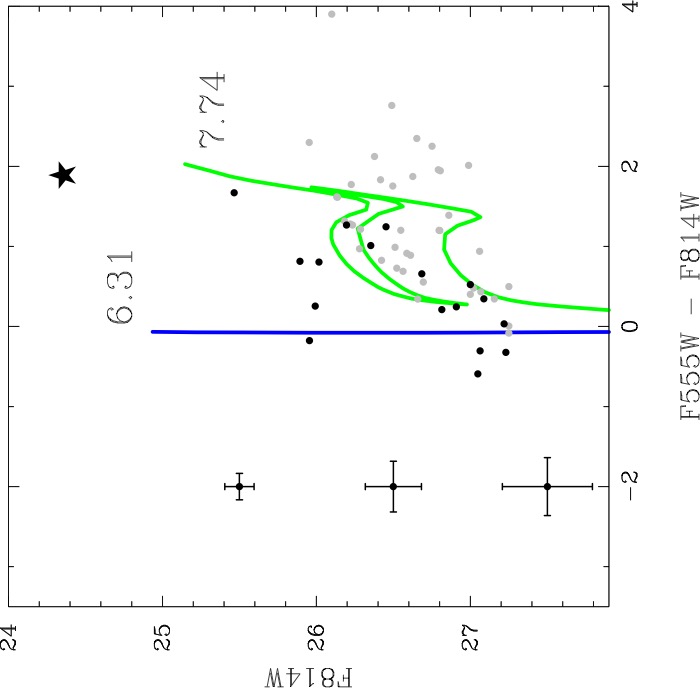}
\vspace{0.2cm}
\end{center}
\caption{Same as for Fig. \ref{fig:cmd:03gd} but for SN 2004A in NGC 6207.}
\label{fig:cmd:04A}
\end{figure}

The stellar population around SN~2004A is shown on the CMD in Fig. \ref{fig:cmd:04A}.  \citet{2014MNRAS.438..938M} noted that the area around SN~2004A was sparsely populated, in part due to the distance to the host galaxy, being one of the furtherest in our sample, as is evident in Fig. \ref{fig:res:img}.  Similar to SN~2003gd, there are a number of bright blue sources consistent with a young stellar population, however the small number of stars (and lack of evolved stars belonging to this population) lead to large uncertainties on the age of the youngest component.  The stellar population around SN~2004A is dominated by an older component, corresponding to the lifetime of a star with initial mass $\sim 7M_{\odot}$.  It is clear from the $F555W-F814W$ CMD of Fig. \ref{fig:cmd:04A} that the position of the progenitor (as determined by \citealt{2014MNRAS.438..938M}) is significantly brighter than would be expected for the age of the older stellar population.   The locus of the progenitor on the CMD, and the mass derived from comparison of the pre-explosion photometry of the progenitor with theoretical SEDs, are consistent with an age of $\sim 25\mathrm{Myr}$ or a star of $M_{init} \sim 10M_{\odot}$.  Such isochrones would not, however, be suitable for the faint, red population that dominates the fit to the older stellar population component.  We conclude, therefore, that the progenitor likely came from an additional third, intermediate age population.  This case highlights the difficulties with determining the properties of a stellar population in instances of low numbers, such that the isochrones are not completely sampled by the observed stars, and the importance of a single star in influencing the interpretation of the population.  We note that the extinction determined for the surrounding stellar population matches the estimates, including the scatter, determined by \citet{2014MNRAS.438..938M}.
%2004dj
%2004dj
%2004dj
\subsection{SN~2004dj}
\begin{figure}
\begin{center}
\includegraphics[width=6.5cm,angle=270]{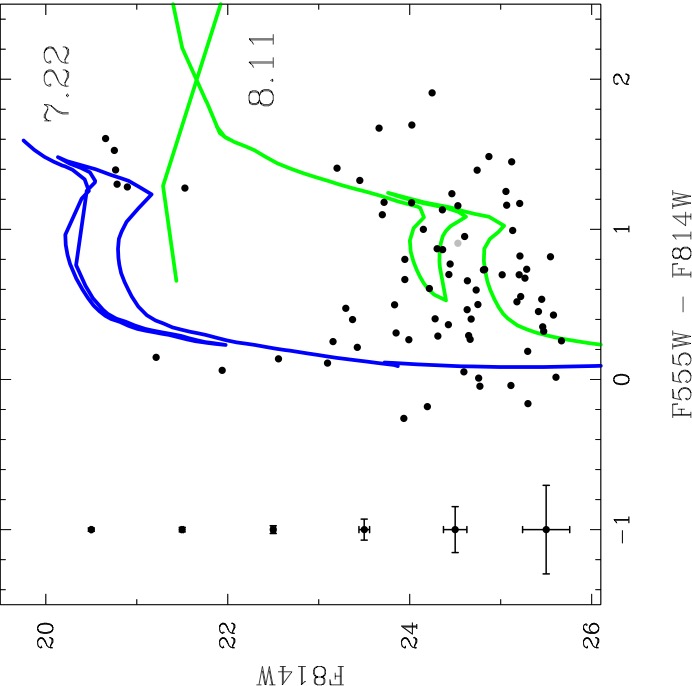}
\vspace{0.2cm}
\end{center}
\caption{Same as for Fig. \ref{fig:cmd:03gd} but for SN 2004dj in NGC 2403.}
\label{fig:cmd:04dj}
\end{figure}

The initial mass for the progenitor of SN~2004dj was constrained through the analysis of the integrated SED of the host cluster Sandage 96 and the surrounding resolved stellar population \citep{2004ApJ...615L.113M, 2005ApJ...626L..89W,2009ApJ...695..619V}.  While \citet{2009ApJ...695..619V} were able to consider just stars within $\mathrm{15\,pc}$, due to the proximity of NGC 2403, in order to conduct a comparable analysis to the other SNe in our sample we consider all stars within a $\mathrm{100\,pc}$ area around Sandage 96.  We exclude, however, those sources identified by {\sc dolphot} as extended or blended in the centre of the cluster.  The CMD for the resolved stars around SN~2004dj is shown in Fig. \ref{fig:cmd:04dj}.    In this one instance, the large scatter around the blue plume suggested that there was significant differential extinction (which is not entirely unexpected due to likely the mixture of field and cluster stars), and so we increased the range for $\mathrm{d}A_{V}$  up to 0.4 mags.  The age determination for this stellar population is relatively robust, especially for the youngest population, where the fit is aided by the presence of $5-6$ bright RSGs.  The age of the young population (16.6Myr) corresponds to an initial mass of $13.0\pm1.4M_{\odot}$.  This age is slightly older than that derived by \citet{2004ApJ...615L.113M} and is at the extreme of the age range determined by \citet{2009ApJ...695..619V} for the resolved stellar population.  Both our derived age and extinction solutions are, however, closer to those determined by \citet{2005ApJ...626L..89W}.  
%2004et
%2004et
%2004et
\subsection{SN~2004et}
\begin{figure}
\begin{center}
\includegraphics[width=6.5cm, angle=270]{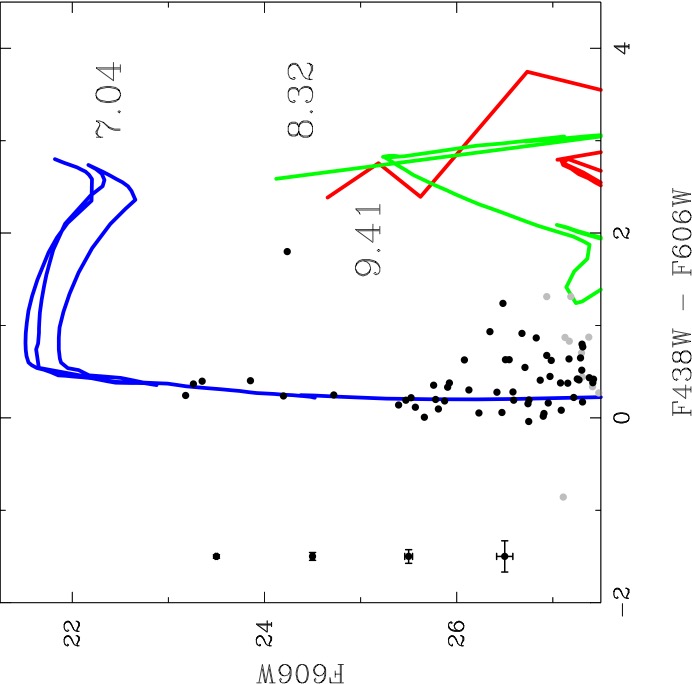}

\vspace{0.5cm}

\includegraphics[width=6.5cm,angle=270]{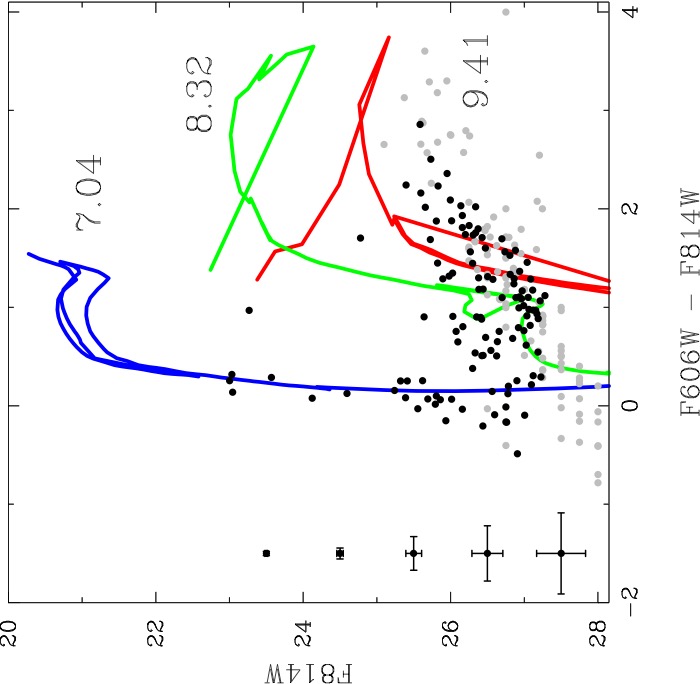}
\vspace{0.2cm}
\end{center}
\caption{Same as for Fig. \ref{fig:cmd:03gd} but for SN 2004et in NGC 6946.}
\label{fig:cmd:04et}
\end{figure}
\citet{2005PASP..117..121L} originally identified a progenitor candidate for SN 2004et in ground-based pre-explosion observations.  While  \citeauthor{2005PASP..117..121L} initially suggested the progenitor was a yellow supergiant, a later analysis of additional pre-explosion observations by \citet{2011MNRAS.410.2767C} showed that the original progenitor candidate was likely to be a blend.  \citeauthor{2011MNRAS.410.2767C} were able to identify a source at the SN position in pre-explosion $I$-band observations, however how much of the brightness of this source could be attributable to the progenitor itself was still uncertain.

In our analysis of the surrounding stellar population (see Fig. \ref{fig:cmd:04et}), we find evidence for three different age components, the youngest of which corresponds to the lifetime of a star with $M_{init} = 17\pm2M_{\odot}$.  The older population components are for stars with $M_{init} < 4M_{\odot}$.

\citeauthor{2011MNRAS.410.2767C} suggest that the progenitor brightness was in the range $21. 27 \leq I \leq 22.06\,\mathrm{mags}$, which would place the progenitor at, or just below, the locus of red supergiants associated with the youngest isochrone.  The mean extinction derived for all stellar populations of $A_{V} = 1.2\,\mathrm{mags}$ is similar to the extinction determined towards SN~2004et itself using interstellar Na absorption lines \citep[$A_{V} = 1.3$ mags;][]{2011MNRAS.410.2767C}.
%2005cs
%2005cs
%2005cs
\subsection{SN~2005cs}
\label{sec:05cs}
\begin{figure}
\begin{center}
\includegraphics[width=6.5cm, angle=270]{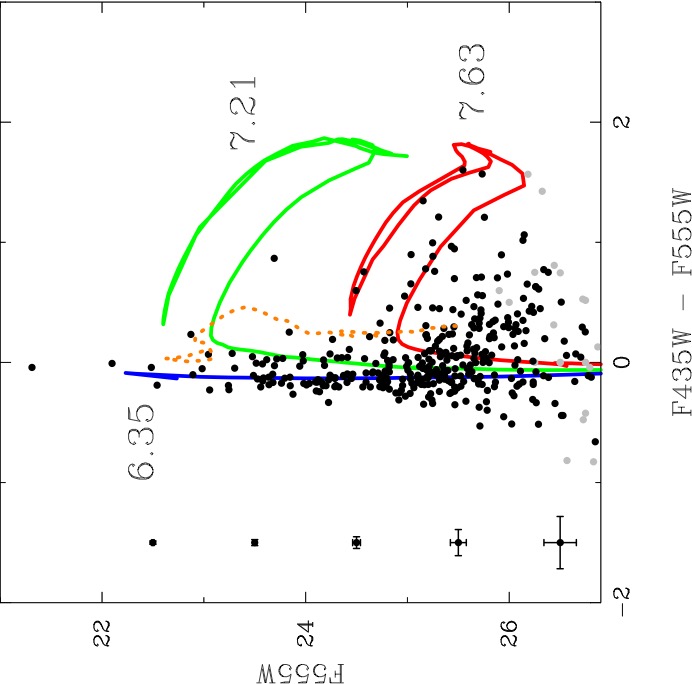}

\vspace{0.5cm}

\includegraphics[width=6.5cm,angle=270]{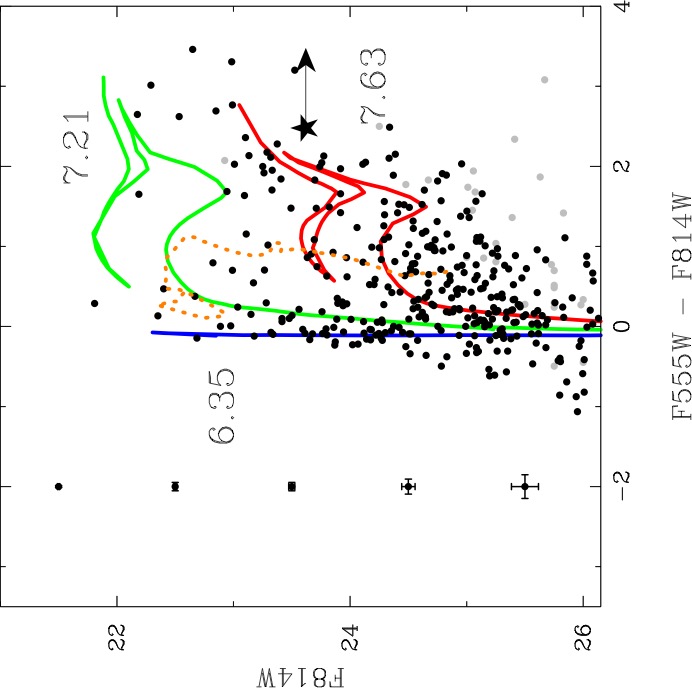}
\vspace{0.2cm}
\end{center}
\caption{Same as for Fig. \ref{fig:cmd:03gd} but for SN 2005cs in M51.The orange dashed line is the colour-magnitude sequence for {\sc starburst99} clusters with mass $10^{3}M_{\odot}$ for ages in the range $1 \leq t \leq 100\,\mathrm{Myr}$, subject to the same extinction as the stellar population.  The position of the progenitor, from the photometry presented by \citet{2014MNRAS.438..938M}, is shown by the black star ($\star$).}
\label{fig:cmd:05cs}
\end{figure}

\citet{2006ApJ...641.1060L} and \citet{2005MNRAS.364L..33M} identified the progenitor of SN~2005cs that occured in the galaxy M51.   Both teams identified a RSG at the SN location in fortuitous pre-explosion HST ACS observations, however the star was only detected in a single $I$-band observation.   A later-analysis by \citet{2014MNRAS.438..938M} showed that this star was indeed the progenitor, having been observed to have disappeared in late-time observations acquired in 2010.   The final mass was estimated to be $\sim 8M_{\odot}$. 
The site of SN~2005cs is quite densely populated and this is reflected on the CMD (see Fig. \ref{fig:cmd:05cs}).  The original progenitor identification was complicated by the proximity of the star to a bright, possibly cluster-like source; so we also consider that there might be contamination by possible clusters, even though we have restricted our sample to those objects that {\sc dolphot} determines to be point-like, unresolved sources.  

In comparison with the expected locus of  {\sc starburst99} model clusters on the CMD (with mass $10^{3}M_{\odot}$), we find that the while there may be some overlap between the brightest individual stars and the oldest clusters (on the $F435W-F555W$ CMD), most young clusters are at the brightest extreme of the sources we observe.  We do not see a separate sequence of sources on the $F555W- F814W$ CMD that might follow an alternative non-stellar track and, hence, conclude that if there is contamination by clusters, it is likely to be very small.  Furthermore, if sources in the CMD are clusters, then the brightness of the brightest sources correspond to cluster masses $< 10^{4}M_{\odot}$, for which it is likely that only a small minority of the stars in those clusters dominate the SED (and, hence, would likely appear to follow stellar isochrones) , rather than the fully populated IMF assumed for {\sc starburst99} models.

We find evidence for three age components on the CMD, whose ages correspond to the lifetimes of stars with masses $M_{init} = 94^{+10}_{-16}$, $13\pm1$ and $7.9\pm0.5M_{\odot}$.  The two older components appear to be blended on the CMD and may actually be a single continuous population, whereas our analysis has assumed them to be composed of discrete age components.  Assuming the progenitor arose from the oldest age component, we find that the corresponding mass matches the initial mass for the progenitor derived from the pre-explosion observations \citep{2014MNRAS.438..938M}.

It is interesting to note that, despite the relatively low mass of the progenitor,  the $F435W - F555W$ CMD is dominated by bright, blue sources corresponding to an age of $\sim 2.2\,\mathrm{Myr}$, with the brightest sources having masses $\sim 50-60M_{\odot}$.  As such, caution is required in inferring the properties of the progenitors from studies of their environments, in that older, low mass progenitors may still be exploding in environments that also contain younger, more massive stars.

%2006ov
%2006ov
%2006ov
\subsection{SN~2006ov}
\begin{figure}
\begin{center}
\includegraphics[width=6.5cm, angle=270]{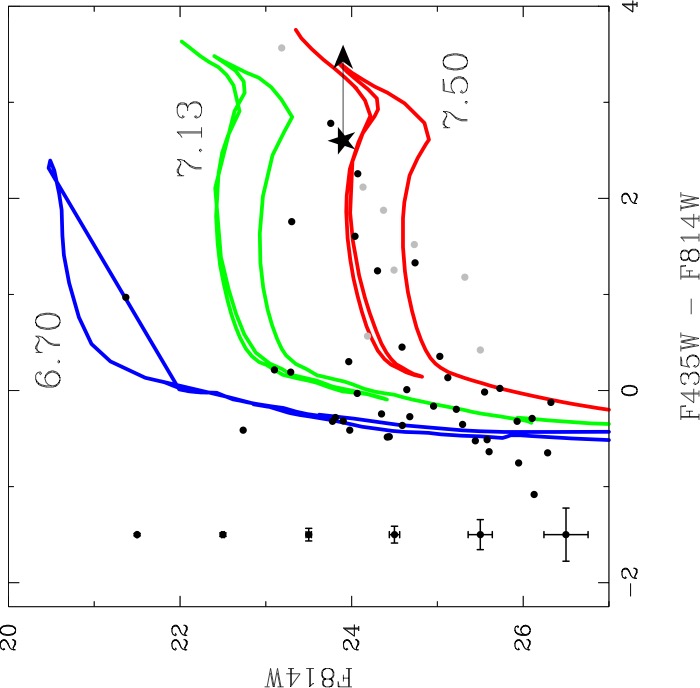}
\vspace{0.2cm}
\end{center}
\caption{Same as for Fig. \ref{fig:cmd:03gd} but for SN 2006ov in NGC 4303.}
\label{fig:cmd:06ov}
\end{figure}
\citet{2007ApJ...661.1013L} originally identified a candidate progenitor of SN~2006ov in NGC~4303, suggesting a relatively high mass, red supergiant with $M_{init} \sim 15M_{\odot}$.  A later study by \citet{2011MNRAS.410.2767C} determined, however, that the pre-explosion candidate and the SN were not spatially coincident, and suggested that the progenitor was not detected; with a limit on its initial mass of $M_{init} \leq 10M_{\odot}$.  The CMD for SN~2006ov (see Fig. \ref{fig:cmd:06ov}) is sparsely population, however the choice of colours ($F435W$ and $F814W$) samples both young and old populations.  We find three population components, although mostly weighted towards the young ($5\,\mathrm{Myr}$, although note the relatively large uncertainties) and old ($31.6\,\mathrm{Myr}$) components, with the latter component being consistent with the life-time of a star with initial mass $9\pm0.6M_{\odot}$.  

\citeauthor{2007ApJ...661.1013L}  assumed only a small degree of Galactic extinction (corresponding to $A_{V} = 0.07\,\mathrm{mag}$), with no correction for the host.  We find a slightly higher level of total extinction ($A_{V} = 0.25\,\mathrm{mag}$), however this is still relatively small (compared to the rest of the sample of Type IIP SNe considered here).  
%2008bk
%2008bk
%2008bk
\subsection{SN~2008bk}
\label{sec:res:08bk}
\begin{figure}
\begin{center}
\includegraphics[width=6.5cm, angle=270]{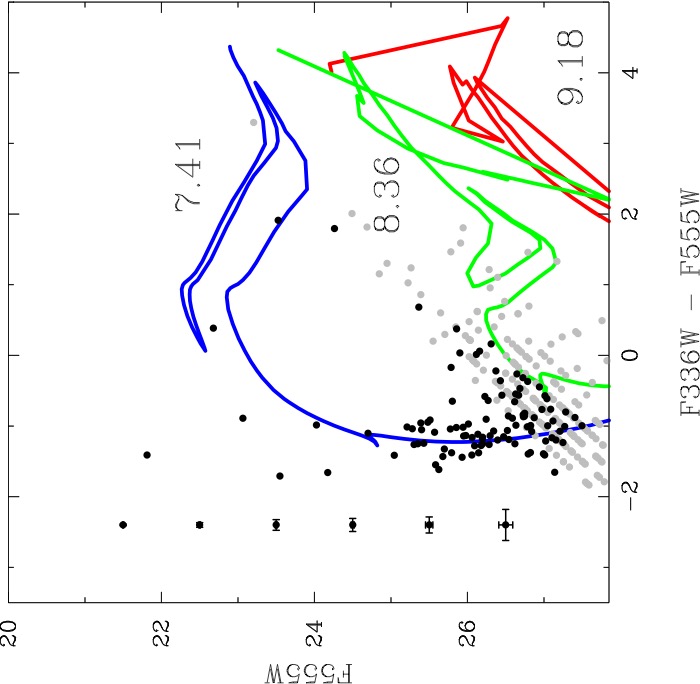}

\vspace{0.5cm}

\includegraphics[width=6.5cm, angle=270]{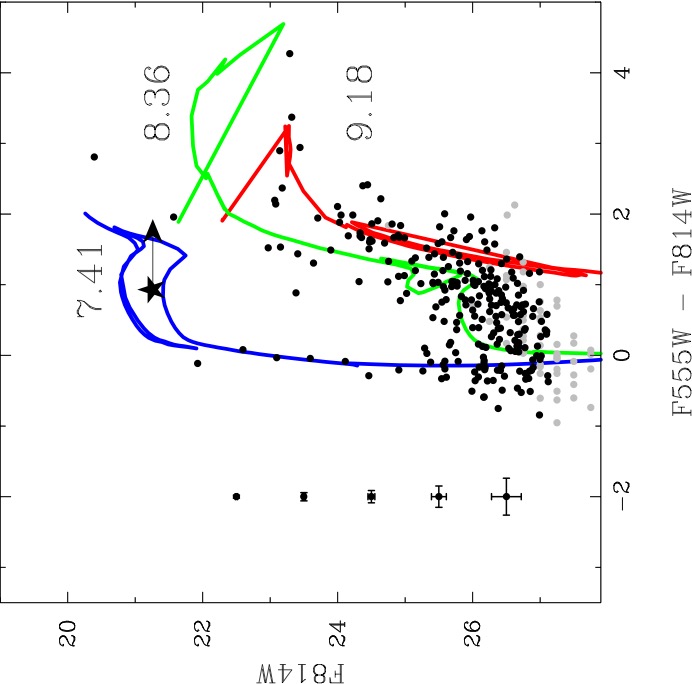}
\vspace{0.2cm}
\end{center}
\caption{Same as for Fig. \ref{fig:cmd:03gd} but for SN 2008bk in NGC~7793.}
\label{fig:cmd:08bk}
\end{figure}

\begin{figure}
\begin{center}
\includegraphics[width=6.5cm, angle=270]{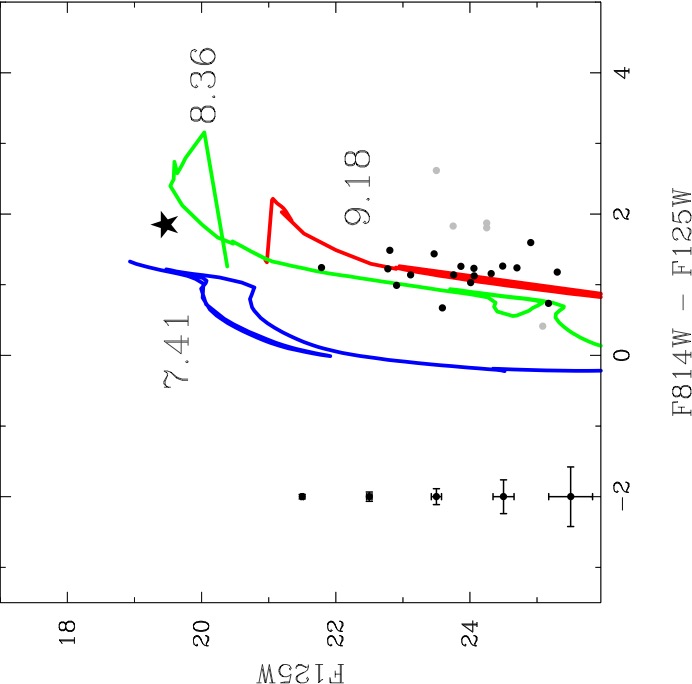}

\vspace{0.5cm}

\includegraphics[width=6.5cm, angle=270]{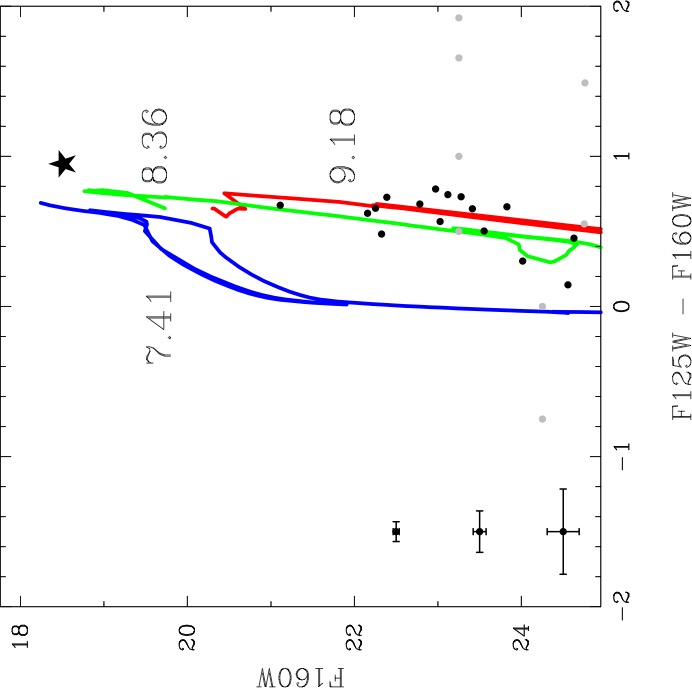}
\vspace{0.2cm}
\end{center}
\caption{HST near-infrared photometry of the stellar population around SN~2008bk.  Overplotted are isochrones corresponding to the fit obtained from just the optical data (see Fig. \ref{fig:cmd:08bk}).}
\label{fig:cmd:08bk:ir}
\end{figure}

\citet{2008ApJ...688L..91M} and \citet{2012AJ....143...19V} reported the detection of a RSG progenitor for SN 2008bk in NGC 7793, in a mixture of optical and near-infrared pre-explosion observations.  Later observations presented by \citet{2014MNRAS.438.1577M} confirmed the disappearance of this star and, using image subtraction techniques, determined a mass of $\sim 13M_{\odot}$; however the progenitor was found to be subject to an extinction of $A_{V} = 2.5\,\mathrm{mag}$.
 
On the CMD (see Figure \ref{fig:cmd:08bk}) we find evidence for multiple age components with the youngest corresponding to an age of $25.7\,\mathrm{Myr}$ or the lifetime of a star with initial mass $11\pm 0.8M_{\odot}$, which is consistent with the high-mass derived by \citet{2014MNRAS.438.1577M} from pre-explosion observations.  We note, however, that the extinction towards the population is significantly lower with $A_{V} = 0.28\,\mathrm{mags}$, which would suggest that the progenitor of SN~2008bk was subject to additional extinction local to that star which did not affect the rest of the surrounding population.  The extinction we derive for the population is also significantly higher than the value  \citet{2012AJ....143...19V} estimated from observations of the SN itself and \citet{2013AJ....146...24V} determined for the subsequent light echo.  

This data highlights the importance of having observations at blue wavelengths, in particular the $F336W$ images, to derive the properties of massive stars, with the shape of young isochrones in the $F336W-F555W$ CMD being less susceptible to the degeneracy between age and extinction than the corresponding $F555W-F814W$ CMD.

Although we have restricted our isochrone fit to the available optical HST data, a useful test of our analysis is to compare our results at different wavelengths.  On Fig. \ref{fig:cmd:08bk:ir}, we  show the results from our analysis of the optical data superimposed on CMDs constructed from near-infrared observations of the same field (including some of the same stars).  Although at these wavelengths we do not probe the young component, both of the older components are present and distinguishable.    We also find that the locus of the progenitor, from the brightness estimated by \citet{2014MNRAS.438.1577M}, is in good agreement with the young isochrones in near-infrared CMD (after considering corrections for the large extinction; although this is reduced at near-infrared wavelengths).  The near-infrared CMD does, however, require some caution in its interpretation, since the Padova models seem to predict similar luminosities for red giants and the more massive RSGs for certain age combinations.

%2009kr
%2009kr
%2009kr
\subsection{SN~2009kr}
\begin{figure}
\begin{center}
\includegraphics[width=6.5cm, angle=270]{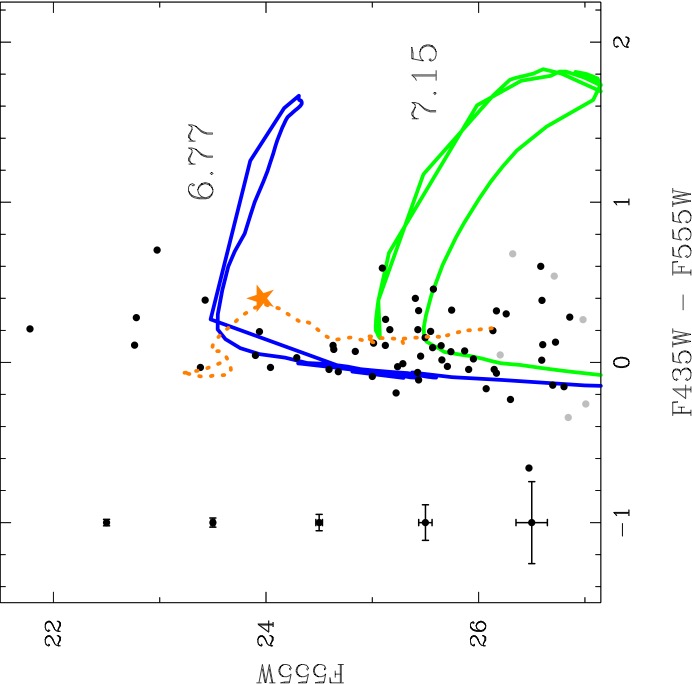}

\vspace{0.5cm}

\includegraphics[width=6.5cm, angle=270]{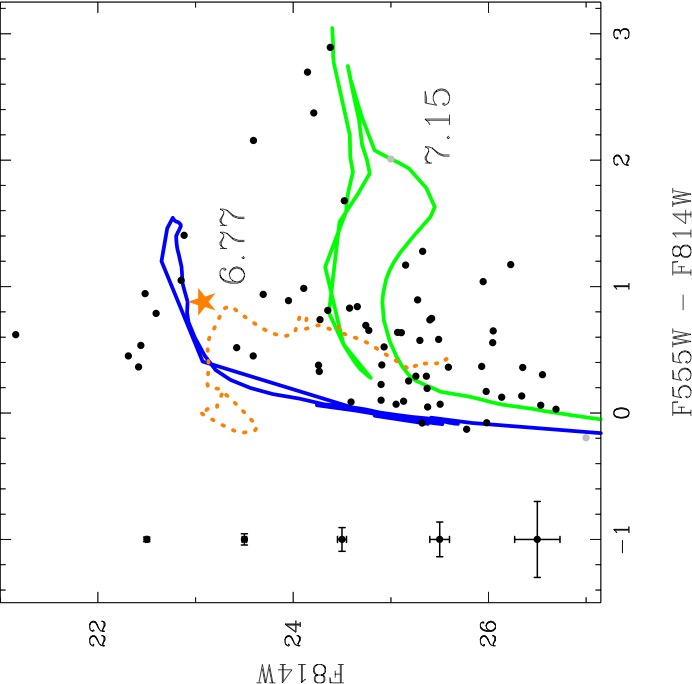}
\vspace{0.2cm}
\end{center}
\caption{Same as for Fig. \ref{fig:cmd:03gd} but for SN 2009kr in NGC 1832.  The orange dashed line is the expect colour-magnitude sequence for $6000M_{\odot}$ compact clusters, from {\sc starburst99}, subject to the same degree of extinction as the stellar population.}
\label{fig:cmd:09kr}
\end{figure}

Using late-time observations, \citet{2015MNRAS.447.3207M} suggested the pre-explosion source detected at the postion of SN~2009kr was likely to be a compact cluster, with mass $\sim 6000M_{\odot}$.  The population of sources around the site of SN~2009kr is shown on the CMD on Fig. \ref{fig:cmd:09kr}, from which it can also be seen that the late-time source at the SN position is consistent with the colour-magnitude sequence for clusters of mass $6000M_{\odot}$.  Given the presence of one source which may be a cluster, and the distance of NGC 1832 prohibiting classification based on spatial criteria,  it is likely that there may be other sources that are also clusters on the CMD.  Using the limit suggested by \citet{2005A&A...443...79B} that sources with $M_{V} < 8.6\,\mathrm{mags}$ are equally likely to be clusters, corresponding to an apparent magnitude limit of $\sim 23.5\,\mathrm{mags}$ we have excluded the brightest 4 sources on the $F435W - F555W$ CMD.  In this case, however, we still cannot authoritatively classify the remaining bright sources as either clusters or bright stars.

Our analysis of the surrounding population suggests the population is composed of two age components.  The extinction we derive toward the population around SN~2009kr, similar to that derived by \citet{2010ApJ...714L.254E} from the strength of interstellar Na absorption lines in the spectrum of SN~2009kr, supports the young age solution for the cluster found by  \citet{2015MNRAS.447.3207M}.  The age of the young stellar population suggests an initial mass for the progenitor of $\sim 30M_{\odot}$, while the older implies a mass of $\sim 15M_{\odot}$.   Although the young stellar component may be contaminated by clusters, there are a number of RSGs evident on the $F555W-F814W$ CMD, clearly supporting this interpretation of the age.
 
The presence of a young stellar population, and possibly a more massive progenitor, is supported by the observations of \citet{2013AJ....146...31K} who estimate an age, from the strength of $\mathrm{H\alpha}$ in integral field spectroscopy, of $3.3\,\mathrm{Myr}$. 

%2009md
%2009md
%2009md
\subsection{SN~2009md}
\begin{figure}
\begin{center}
\includegraphics[width=6.5cm, angle=270]{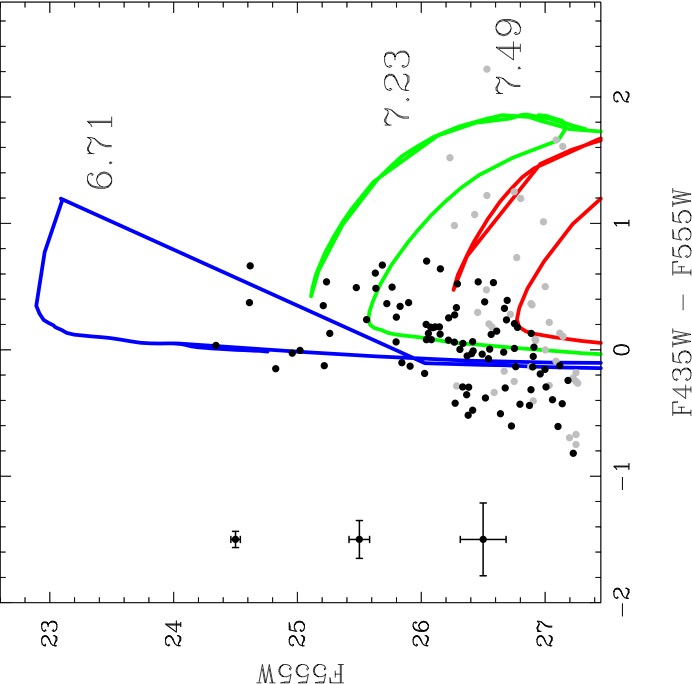}

\vspace{0.5cm}

\includegraphics[width=6.5cm, angle=270]{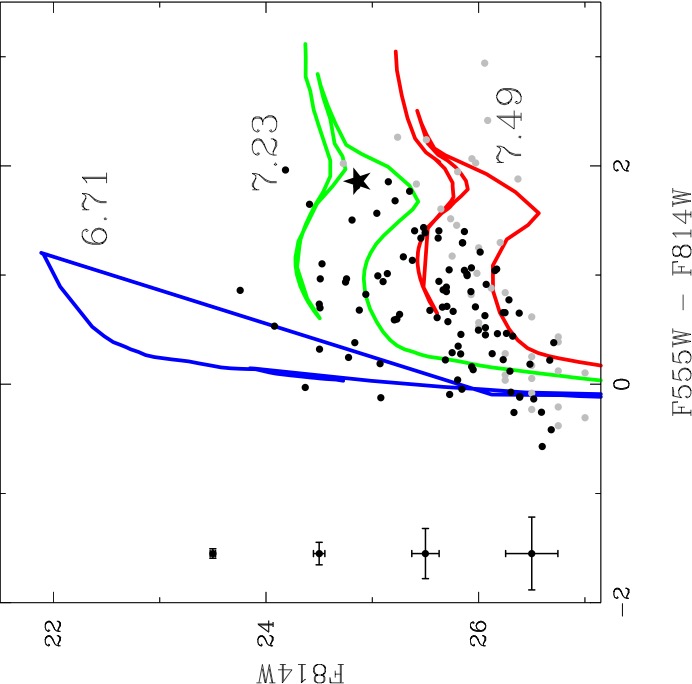}
\vspace{0.2cm}
\end{center}
\caption{Same as for Fig. \ref{fig:cmd:03gd} but for SN 2009md in NGC 3389.}
\label{fig:cmd:09md}
\end{figure}
The surrounding stellar population for SN 2009md is shown on the CMD on Figure \ref{fig:cmd:09md}.  We find that the stellar population can be described by three age components.  The extinction we find for the surrounding stars ($A_{V} = 0.55\,\mathrm{mags}$) is consistent with the reddening ($E(B-V) = 0.2\,\mathrm{mags}$) adopted by \citet{2015MNRAS.447.3207M} for the analysis of the progenitor itself.
Using the values for the brightness of the progenitor candidate and the source recovered at the SN position in late-time observations by \citeauthor{2015MNRAS.447.3207M}, we find that the brightness of the pre-explosion source is  consistent with the $17\,\mathrm{Myr}$ isochrone  or the lifetime of a star with $M_{init} = 13\pm1M_{\odot}$.  This mass estimate is consistent with the constraints placed by \citeauthor{2015MNRAS.447.3207M} on the progenitor of $8.5-13M_{\odot}$ (where the large mass range reflects their uncertainties on the reddening).

%2012aw
%2012aw
%2012aw
\subsection{SN~2012aw}
\label{sec:12aw}

\begin{figure}
\begin{center}
\includegraphics[width=6.5cm, angle=270]{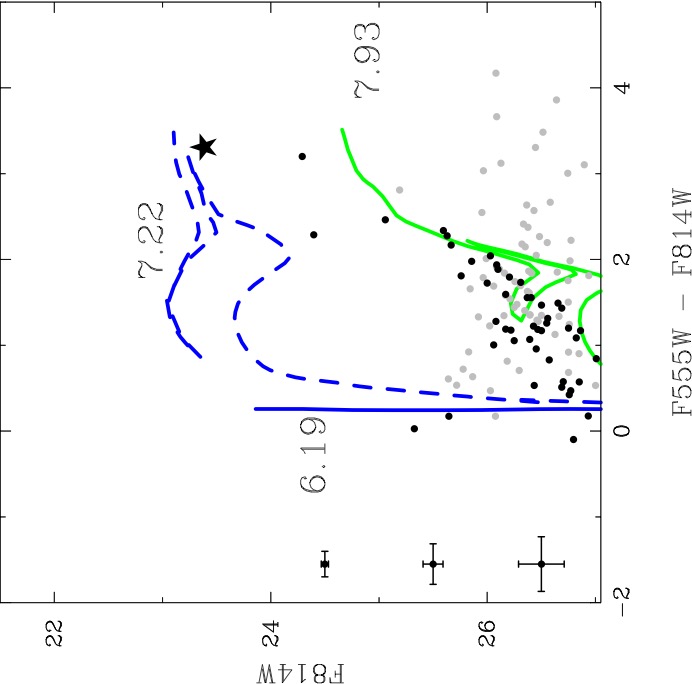}
\vspace{0.2cm}
\end{center}
\caption{Same as for Fig. \ref{fig:cmd:03gd} but for SN 2012aw in NGC 3351.  The dashed isochrone corresponds to the fit including the pre-explosion photometry of the progenitor.}
\label{fig:cmd:12aw}
\end{figure}

The progenitor of SN~2012aw was identified in pre-explosion optical
and near-infrared observations by \citet{2012arXiv1204.1523F} and \citet{2012ApJ...756..131V}, although difficulties with the
determination of the reddening towards the progenitor (as opposed to
the reddening determined towards the SN itself) yielded large uncertainties on the progenitor's initial mass. \citet{2012ApJ...759...20K},
using theoretical models of dust, suggested that the progenitor may
have been subject to reddening arising in the circumstellar medium
with a different dust composition to normal interstellar dust.

The field around SN~2012aw is notable for the presence of multiple
sources found in the $F814W$ observation, but a dearth of
sources that are also found in the corresponding $F555W$
image.  On a qualitative level, this already suggests that
significant reddening may be present.  We find the population to be composed of two age components: a young component, primarily influenced by $\lesssim 5$ stars which are detected in the $F555W$ image, and an older population, requiring a high
a degree of extinction $A_{V} = 1.37\,\mathrm{mags}$ (see Fig. \ref{fig:cmd:12aw}).  This extinction
supports the high-extinction interpretation of the pre-explosion
observations presented in \citet{2012arXiv1204.1523F}, implying a progenitor
mass of $12-14M_{\odot}$.  We note, however, that this value for the
extinction is significantly higher than the extinction that \citet{2015ApJ...806..195V}
 require for the production of the light echo ($A_{V} =0.24\,\mathrm{mags}$).

\begin{figure}
\begin{center}
\includegraphics[width=6.5cm,angle=270]{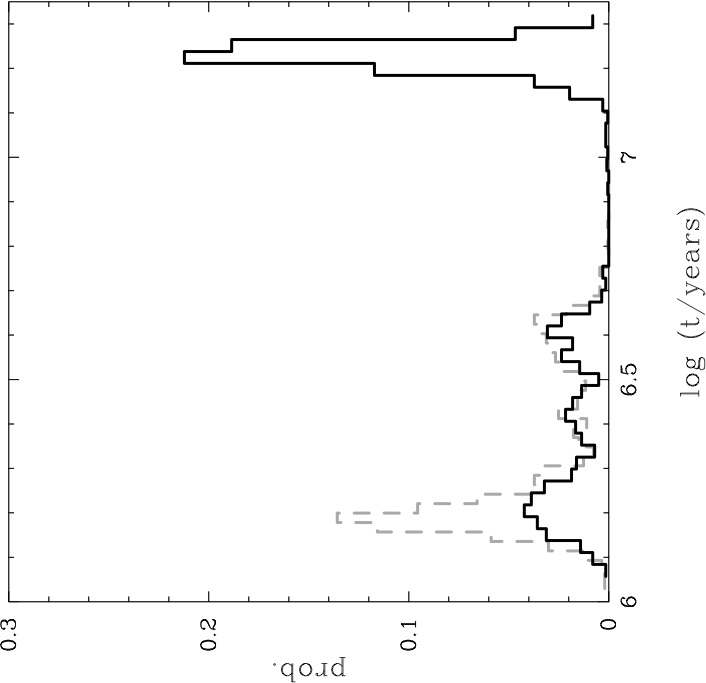}
\vspace{0.2cm}
\end{center}
\caption{Posterior probability distributions for the age of the young stellar population component around SN~2012aw; with (solid black line) and without (dashed grey line) the progenitor included in the fit.}
\label{fig:12aw:comp}
\end{figure}

Similarly to SN~2004A (see Section \ref{sec:04A}), the age determined for the young stellar component is very uncertain, in part because of the small number of stars that are identified as belonging to it (see Table \ref{tab:res}). 
Because the progenitor of SN~2012aw was detected in the same two photometric bands used in the analysis of the surrounding population, we also conducted an iteration of our analysis with the progenitor included. We find that the properties of the fit
remain constant, with two age components at the same extinction, except that the posterior probability density
function  for the age of the young population component has an
additional mode at $16.6\,\mathrm{Myr}$ (see Fig. \ref{fig:12aw:comp}) corresponding to the lifetime
of a star with  $M_{init} = 13.5\pm1M_{\odot}$ (this fit is also
shown on Fig. \ref{fig:cmd:12aw}).  It is important to note that this does not constitute an additional stellar population component, but rather an alternative interpretation of the photometry of the small number of stars that make up the young component.  The derived mass for this age component is consistent with the range derived from pre-explosion images \citep{2012arXiv1204.1523F} and from models of the late-time SN spectrum \citep{2014MNRAS.439.3694J}.

In the absence of the progenitor detection, and assuming the
progenitor arose from the old population component, we would consider the
age of $85\,\mathrm{Myr}$ to correspond to the lifetime of a $\sim
6M_{\odot}$ star (which would represent a significant underestimate).

\begin{figure}
\begin{center}
\includegraphics[width=6.5cm, angle=270]{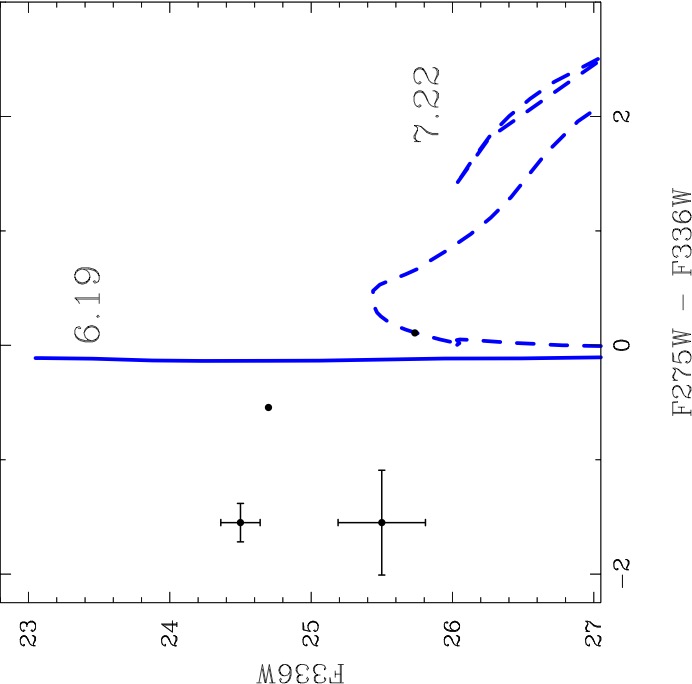}
\vspace{0.2cm}
\end{center}
\caption{Same as for Fig. \ref{fig:cmd:03gd} but for LEGUS $F275W$ and $F336W$ observations of the site of SN 2012aw in NGC 3351.  The dashed isochrone corresponds to the fit including the pre-explosion photometry of the progenitor.}
\label{fig:cmd:12aw:uv}
\end{figure}

We note that HST observations of the site of SN~2012aw were also acquired
as part of the ``Legacy ExtraGalactic UV Survey" (LEGUS; GO-13364; PI Calzetti) on 2014
April 23, and cover a larger wavelength range than the dataset we
have used here, although they  are
significantly shallower than our observations (with the exposure time in
$F555W$ a factor of 3 smaller than our dataset).  We considered the UV observations acquired with the $F275W$ (2361s) and $F336W$ (1062s) filters, as shown on Fig. \ref{fig:cmd:12aw:uv}.  Within the same search selection radius, only two stars are recovered at UV wavelengths, supporting the high level of extinction inferred from the optical observations.  The photometry of these two stars could be consistent both with the young and intermediate stellar population ages identified in the optical CMD.
%2012ec
%2012ec
%2012ec
\subsection{SN~2012ec}
\label{sec:12ec}
\begin{figure}
\begin{center}
\includegraphics[width=6.5cm, angle=270]{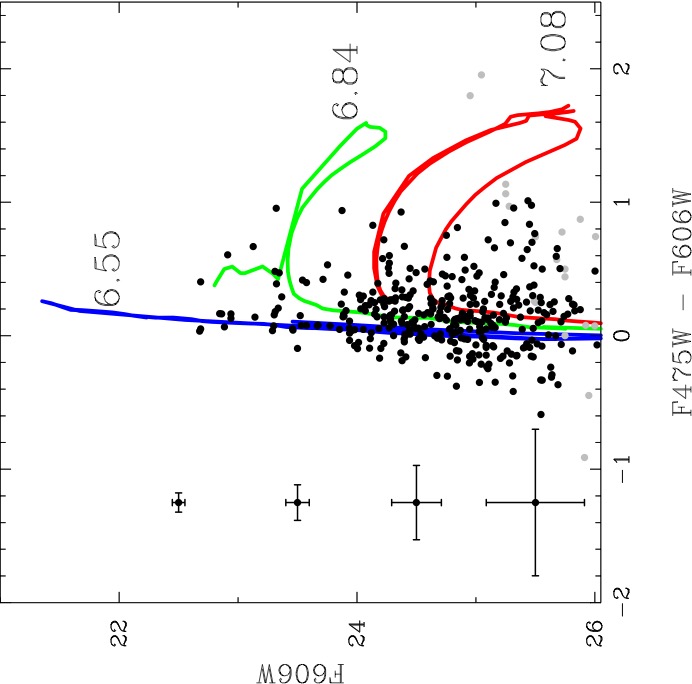}

\vspace{0.5cm}

\includegraphics[width=6.5cm, angle=270]{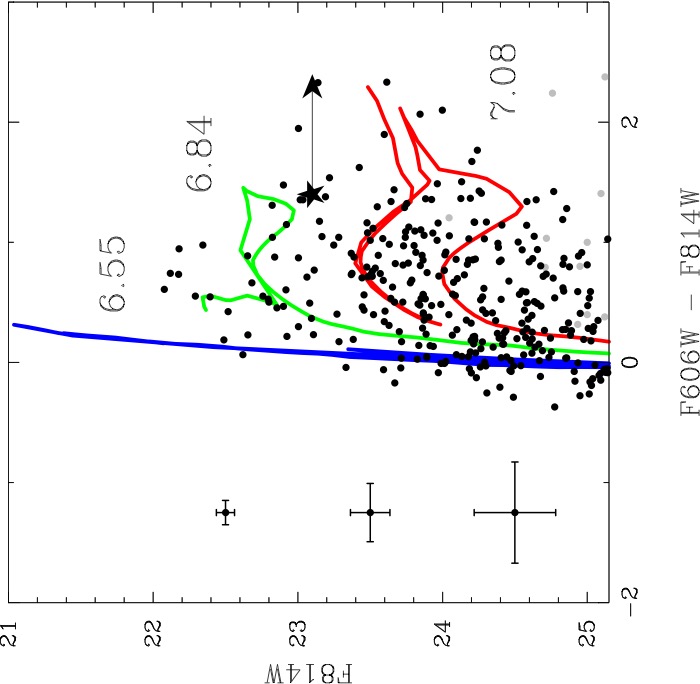}
\vspace{0.2cm}
\end{center}
\caption{Same as for Fig. \ref{fig:cmd:03gd} but for SN~2012ec in NGC~1084.}
\label{fig:cmd:12ec}
\end{figure}

\citet{2013arXiv1302.0170M} determined a mass of $14-22M_{\odot}$ for the RSG progenitor candidate identified in pre-explosion images of the site of SN~2012ec.  The large range in mass was due to the significant uncertainties on the reddening (and corresponding temperature) and the degree of blending/crowding at and around the progenitor location.  To study the surrounding stellar population, we have utilised  recent three-colour observations of the site of SN~2012ec from 2016. Similarly to SN~2005cs (see section \ref{sec:05cs}), we find evidence on the CMD (see Fig. \ref{fig:cmd:12ec}) that the surrounding stellar population may possibly be composed of a continuous range of stellar ages.  \citeauthor{2013arXiv1302.0170M} estimated a reddening of $E(B-V) = 0.10^{+0.15}_{-0.02}\,\mathrm{mags}$ from the strength of interstellar Na absorption lines (using the Poznanski et al. 2012 relation).  Our value for the extinction, derived from the surrounding stellar population, is consistent with the upper $1\sigma$ uncertainty, but also includes significant differential extinction (similar to the value determined for the population in the vicinity of SN~2012aw).  This extinction value, however, would support a higher mass progenitor from the mass range determined by \citeauthor{2013arXiv1302.0170M}.  There is an obvious population of blue sources, although considering their brightness and colours, similar to arguments made concerning the population around SN~2005cs, we find no evidence that there is significant contamination from clusters.  Plotting the progenitor on the CMD, we find it lies intermediate between the two older age components, implying an age of between $6.9$ to $12\,\mathrm{Myr}$ corresponding to the lifetimes of stars with masses in the range $16 - 27M_{\odot}$.  

%2013ej
%2013ej
%2013ej
\subsection{SN~2013ej}
\label{sec:13ej}
\begin{figure}
\begin{center}
\includegraphics[width=6.5cm, angle=270]{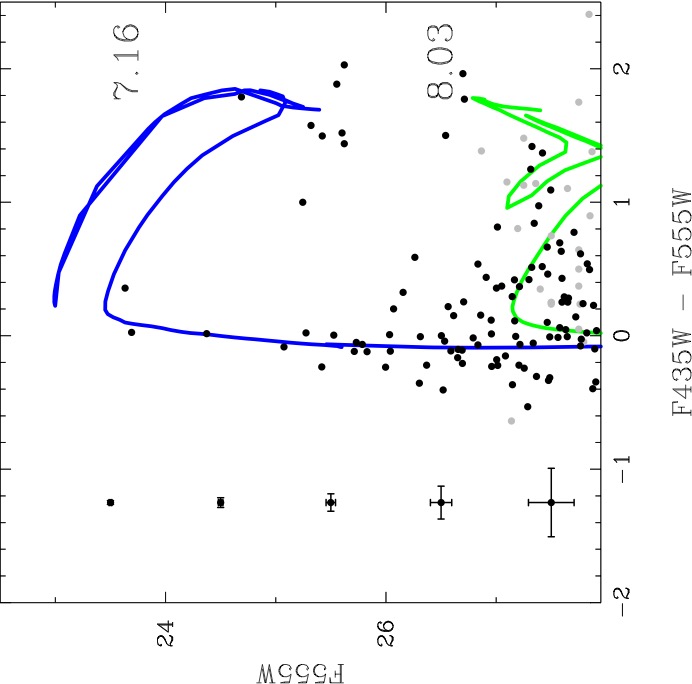}

\vspace{0.5cm}
\includegraphics[width=6.5cm, angle=270]{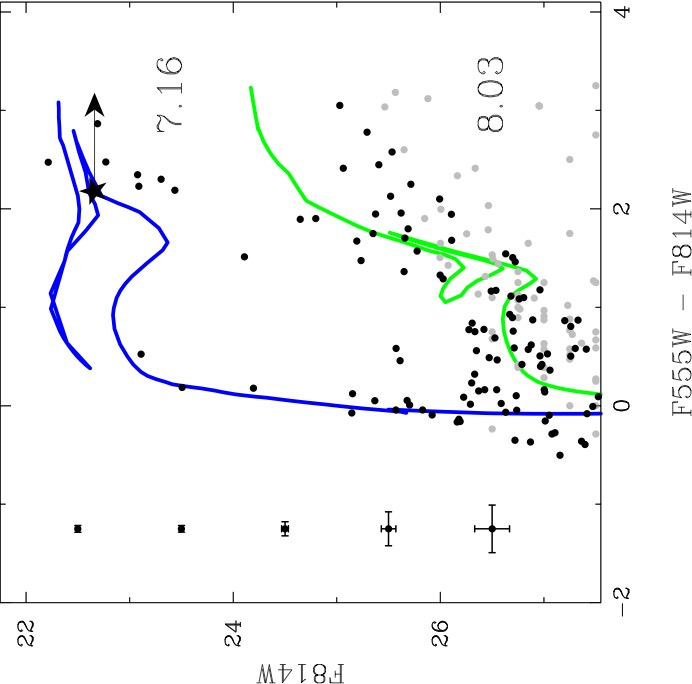}
\vspace{0.2cm}
\end{center}
\caption{Same as for Fig. \ref{fig:cmd:03gd} but for SN 2013ej in M74.}
\label{fig:cmd:13ej}
\end{figure}
The progenitor of SN2013ej was identified in pre-explosion HST ACS observations, the same used for our population analysis, although the source was blended with a neighbouring blue source that prohibited an accurate determination of the colour \citep{2014MNRAS.439L..56F}.
It is obvious from the CMDs (see Fig. \ref{fig:cmd:13ej}) that the population is relatively well-structured, with little scatter about the isochrone on the blue plume, and a well defined population of RSGs with similar brightness to the progenitor candidate.  This supports our finding of low differential extinction. 
While \citet{2014MNRAS.439L..56F} assumed that the progenitor was only subject to the foreground extinction of $A_{V} = 0.19\,\mathrm{mag}$, our analysis of the population suggests there is significant extinction in the host. The age determined for the young stellar population ($14.5\,\mathrm{Myr}$) supports a mass for the progenitor of $14\pm1.5M_{\odot}$, at the upper end of the range proposed by \citeauthor{2014MNRAS.439L..56F} and agrees with the subsequent analysis by \citet{2016MNRAS.461.2003Y}.  
%%%%%%%%%%%%%%%%%%%%%%%%%%%%%%%%
%DISCUSSION
%%%%%%%%%%%%%%%%%%%%%%%%%%%%%%%%
\section{Discussion}
\label{sec:disc}

\begin{figure*}
\includegraphics[width=11cm]{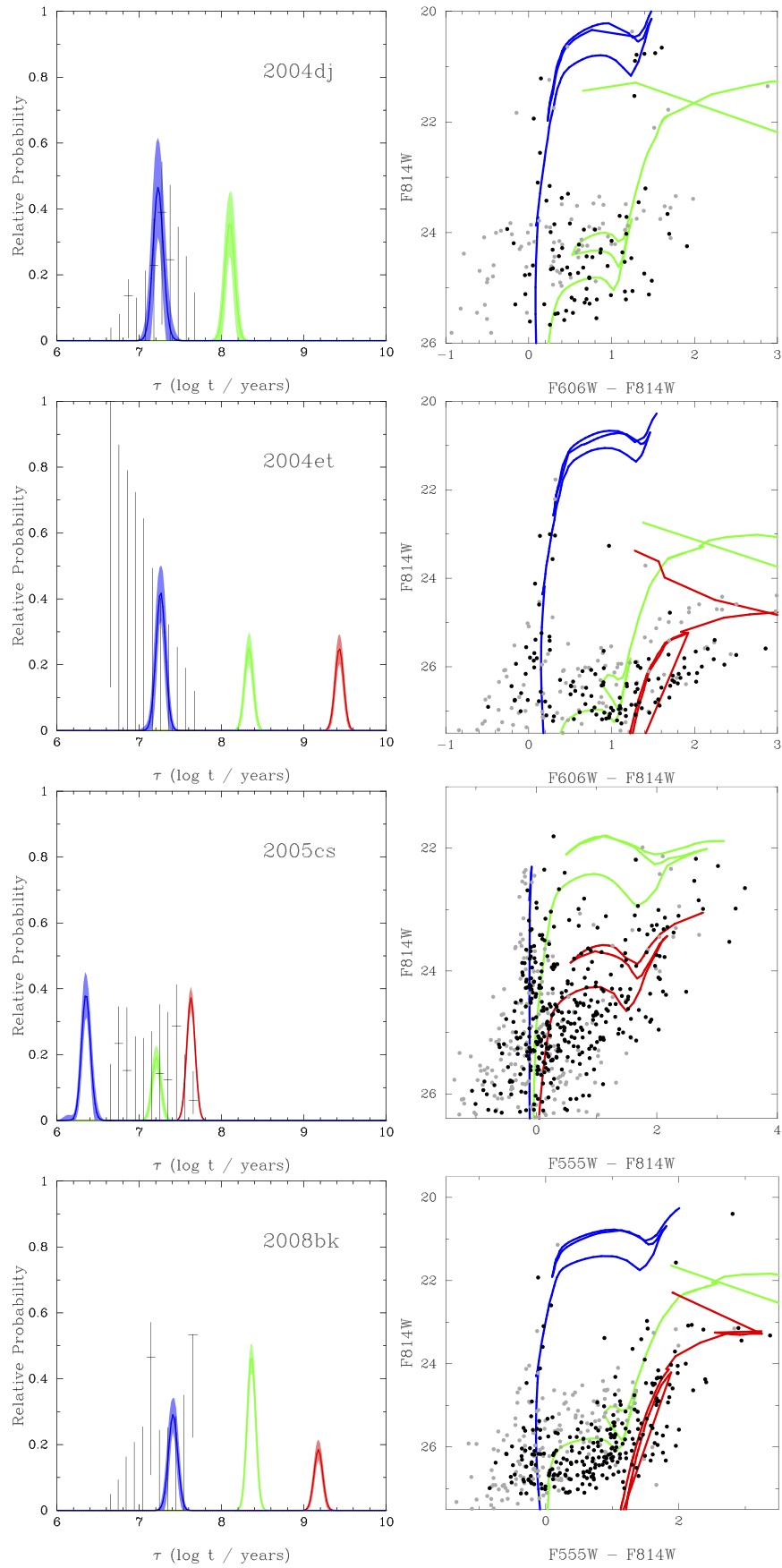}
\caption{{\it Left)} Comparison between the recovered probabilities for the populations around four Type IIP SNe from the Bayesian mixture model (coloured, shaded curves) and the progenitor mass probability function (solid lines), as a function of population mass, derived by \citet{2014ApJ...791..105W}. {\it Right)} A qualitative comparison between the observed stellar population on the CMD (black points) against a synthetic population, generated using the results from the Bayesian mixture model (light grey).  Note that the synthetic population represents merely one realization of a possible CMD observable given the derived population parameters and is subject to Poisson noise. }
\label{fig:disc:compplot}
\end{figure*}

Previously, \citet{2014ApJ...791..105W} analysed the surrounding stellar populations of five of the SNe contained in our sample: 2003gd, 2004dj, 2004et, 2005cs and 2008bk.  Of these, \citeauthor{2014ApJ...791..105W} could not determine a likely mass for the progenitor of SN~2003gd due to the paucity of stars in the vicinity of the SN (of which they only find four)\footnote{We note that we are unable to identify which specific datasets \citeauthor{2014ApJ...791..105W} used in their analysis.}.  In Figure \ref{fig:disc:compplot} we compare the results of our analysis, with the probability estimate for the ages of the progenitors derived by \citet{2014ApJ...791..105W}.  We have used the Padova isochrones for the appropriate metallicity to convert their probability bins defined in units of mass to units of progenitor lifetime in $\tau$.  We note that \citeauthor{2014ApJ...791..105W} only computed the progenitor mass probability functions down to masses corresponding to a maximum progenitor lifetime of $50\,\mathrm{Myr}$.   In all cases, but especially for SN~2004dj, there is agreement between the analysis of \citeauthor{2014ApJ...791..105W} and where we find at least one of our stellar population age components.  The larger uncertainties of \citeauthor{2014ApJ...791..105W}, however, lead to a larger range in possible progenitor masses than we find; this is consequence of the different approaches of the two techniques, where \citeauthor{2014ApJ...791..105W} attempt to the recover the star formation history as the superposition of continuous star formation of varying strengths, we have assumed discrete star formation episodes with fixed temporal width.  

In Figure \ref{fig:disc:compplot}, we also show for a qualitative comparison example synthetic CMDs, constructed from the population histories derived here and generated using Monte Carlo procedures, compared to the observed CMDs.  The Monte Carlo procedure includes the generation of binary systems, the range of ages of each population, the relative number weight of each population and the degree of differential extinction.  We convolved the photometry of our synthetic stars with the uncertainty as a function of magnitude derived from artificial star tests.  We have assumed a single detection limit to be applicable to all the stars in each filter.  In general, we find the loci of synthetic stars and observed stars to overlap extremely well, given Poisson noise.  For SN~2004dj, we note that there are more observed RSGs than expected from the model and the width of the observed blue plume is narrower than the prediction (possibly suggesting the synthetic photometric uncertainties are overestimated).

\citet{2013AJ....146...31K} used integral field spectroscopy of the sites of nearby SNe to determine the ages of the progenitors, using the strengths of $\mathrm{H\alpha}$ and the Ca\,{\sc ii} triplet; and their sample contains three of the SNe considered here.   The limit they derive for SN~2008bk ($>15\,\mathrm{Myr}$) is consistent with the age we have determined, however the usefulness of that measurement is limited as it applies to two clusters offset from the SN position by $120$ and $180\,\mathrm{pc}$.  The limitations of their technique when applied to sources at large offset distances from the SN location is highlighted by SN~2009kr, where the nearest $\mathrm{H\alpha}$-bright source is located $170\,\mathrm{pc}$ from the SN position and yields a significantly younger age ($3.2\,\mathrm{Myr}$) than is evident from our observations of the resolved stellar population.  For SN~2004dj, we find agreement with the age estimated by \citeauthor{2013AJ....146...31K} for the parent cluster (although this is discrepant with the age they measure for an offset {\sc H ii} region).    

A key limitation in studying the resolved stellar populations around SNe is distance, which can limit the area in which individual stars might be observed and, for similar exposure times, limit detections to only a small number of the brightest stars in the field.   The effects of distance are closely coupled with crowding,  which makes identifying and assigning photometry to individual stars difficult along with the commensurate increase in photometric uncertainties.  The issue of crowding is highlighted by the case of SN~2012ec, which has the largest number of surrounding stars in the sample, despite being one of the most distant SNe.  As suggested in Section \ref{sec:12ec}, the large scatter in the positions of stars in the CMDs may reflect continuous star formation, rather than the episodic star formation assumed in the fits; however the photometric uncertainties are also large due to the degree of crowding which means the apparent spread in the brightness and colours of stars may actually be due to limitations in the ability to conduct photometry of individual sources.  At the same time, the inclination of the host galaxy of SN~2012ec is relatively large ($i = 50^{\circ}$; Quoted from \footnote{http://leda.univ-lyon1.fr/}) meaning that the effective area from which we have extracted photometry maybe larger than for a face-on spiral galaxy.  In addition, the applicability of this technique to determine the properties of the progenitors of SNe presupposes that the SN occurred in the natal region of the progenitor.  There is, however, evidence in the Galaxy for a population of runaway OB stars, ejected by gravitational interactions in dense cluster \citep[e.g.][]{2011Sci...334.1380F} or through the explosion of a binary companion \citep[e.g.][]{1997ApJ...475L..37K}, that might evolve to become RSGs with high space velocities \citep{2012ApJ...751L..10M}.   Given the velocity of Betelgeuse of $50\,\mathrm{km\,s^{-1}}$  \citep{2012ApJ...751L..10M}, which is similar to the maximum mean escape velocities for stars from clusters \citep{2011Sci...334.1380F}, over the total lifetime (44Myr) an $8M_{\odot}$ star might travel $\approx 2000\,\mathrm{pc}$.   Taking into account the relatively small fraction of runaway massive stars \citep[20\%;][]{1986ApJS...61..419G}, the effects of projection (i.e. that some portion of the motion will be in a radial component) and the time at which the star becomes a runaway, only a small number of supernovae might be expected to have such an extreme offset from the site of their progenitor's birth.

\begin{figure}
\includegraphics[width=7.5cm,angle=270]{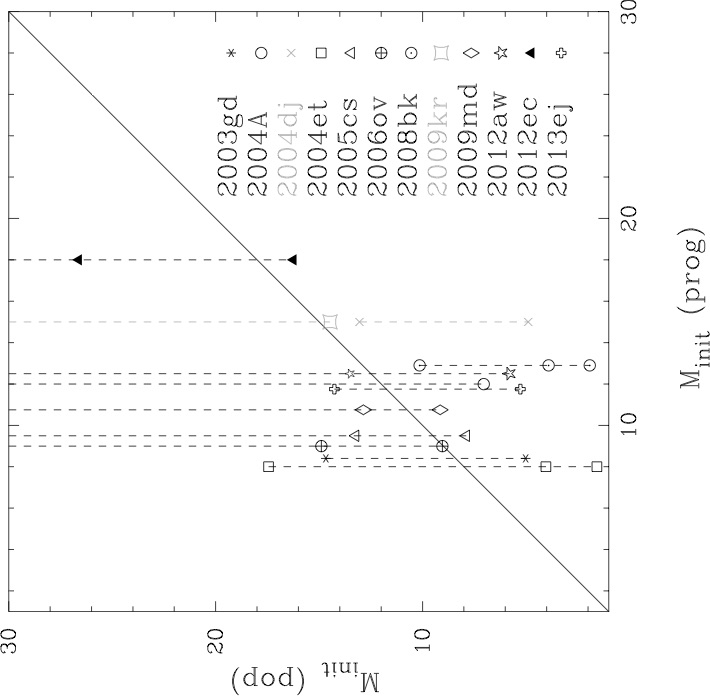}
\caption{Comparison between previously reported progenitor masses derived from direct imaging of the progenitors in pre-explosion observations (for references see Table \ref{tab:obs:samp}) and the masses of stars with lifetimes derived from analysis of the stellar populations (for all age components) presented here.}
\label{fig:disc:mass}
\end{figure}

In Figure \ref{fig:disc:mass} we compare the masses derived from the ages of the stellar populations (see Figure \ref{fig:res:massage}), for each of the age components where applicable, with those derived for the progenitors observed in  pre-explosion images.  In the case of this sample of Type IIP SNe, we find general agreement to within $\pm 3M_{\odot}$ between the masses determined through the two approaches, assuming the progenitor arose from the nearest age component appropriate for the lifetimes of RSG progenitors.   Some caution is required in this interpretation, because in our comparison we have implicitly assumed that the pre-explosion observations provide the gold-standard mass determination, which, in of itself, is not necessarily the case as these may also be subject to significant uncertainty.   An example is SN~2004et, where the ambiguous interpretation of the photometry of the pre-explosion observations \citep[and the unknown spectral type of the progenitor;][]{2011MNRAS.410.2767C} means that the mass derived from direct imaging may be uncertain.  The pre-explosion photometry of the progenitor candidate may be consistent with the youngest isochrone and, therefore, a much higher mass ($\sim 17M_{\odot}$).   This higher mass is consistent the mass estimated from the strength of O emission from late-time observations of the SN \citep{2012A&A...546A..28J}, although not as high as the mass derived from hydrodynamical modelling \citep{2009A&A...506..829U}. 

Reassuringly, the age estimates for the resolved populations surrounding SNe 2004dj and 2009kr are in agreement with the analyses of the integrated SEDs of the clusters found at the positions of these SNe in pre-explosion images. In two cases, SNe~2004A and 2012aw, there are discrepancies that arise because of the small number of stars; but such circumstances are readily identifiable by the probability density function for the age of that particular age component exhibiting multiple modes (see Fig. \ref{fig:12aw:comp}).  
To be truly useful for the systematic determination of progenitor masses, this technique requires some way of identifying securely which population the progenitor actually arose from.  In some cases, such as for SN~2008bk, only one of the population components is young enough to still contain stars massive enough to explode as SNe; however, this is not true for half of the SNe in our sample.  Of those, the progenitors of four SNe are not consistent with the youngest age component.  For SN~2005cs, there are stars from the youngest age component with masses of $\sim 50 - 60M_{\odot}$, which refutes the claim by \citet{2016arXiv161109264M} that there is an ``the absence in colour magnitude diagrams of stars in the environment of historic core-collapse supernovae of stars with $>20M_{\odot}$".

\begin{figure}
\includegraphics[width=7.5cm,angle=270]{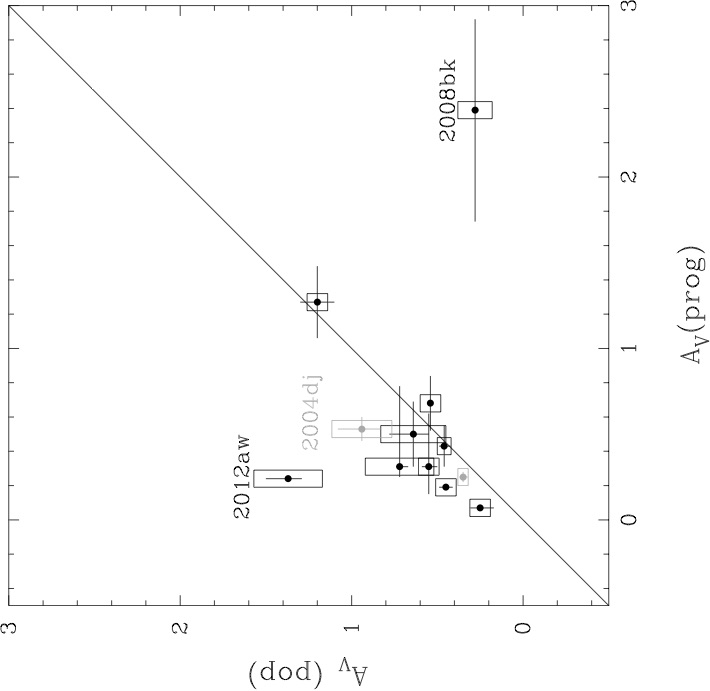}
\caption{Comparison of the extinction estimates used in the analysis of detected progenitor candidates (for references see Table \ref{tab:obs:samp}) and extinctions derived from our analysis of the surrounding population.  Error bars are shown for values of $A_{V}$, while boxes indicate the degree of differential extinction $\mathrm{d}A_{V}$ derived from the analysis of the population only.  Objects indicated in grey, associated with SNe 2004dj and 2009kr, are candidate host clusters for the progenitor and not individual stars.}
\label{fig:disc:ext}
\end{figure}

We find the extinctions determined for the stellar populations are on average $\sim 0.3\,\mathrm{mags}$ larger than those values {\it applied} to the progenitors in the analysis of the pre-explosion images, although there are some extreme outliers like SNe 2008bk and 2012aw (see Figure \ref{fig:disc:ext}).  For the analysis of the progenitor objects, a range of different approaches are used to constrain the extinction.  For example, both \citet{2014MNRAS.438.1577M} and \citet{2012AJ....143...19V} used fits to the SED of the photometry of the progenitor to constrain the reddening (and, because of different interpretation of the data, arrived at different conclusions), while \citet{2014MNRAS.438..938M} used SED fits to surrounding stars to estimate the Galactic and host reddening to the progenitor of SN~2003gd and \citet{2014MNRAS.439L..56F} assumed only Galactic reddening, from dust maps \citep{2011ApJ...737..103S}, towards the progenitor of SN~2013ej.   An alternative methodology, as used for a SN in our sample by \citet{2013arXiv1302.0170M} for example, is to utilise observations of interstellar Na lines in spectra of the subsequent SN and correlate this with a degree of reddening \citep[][]{2003fthp.conf..200T,2012MNRAS.426.1465P} .   It is important to note that, apart from SED fits to the progenitor itself, all of the techniques for estimating the reddening only probe the Galactic and host dust, but will exclude dust that may be local to the progenitor and destroyed by the SN \citep{1986MNRAS.221..789G} and will not be seen in post-explosion observations.

As such the locus of points on Figure \ref{fig:disc:ext} does not refer to the extinction measured to the progenitor, but rather the extinction {\it assumed} to affect the progenitor's pre-explosion photometry.  This discrepancy between the population and progenitor extinctions is evident for SN~2008bk, where there is significant evidence for circumstellar dust around the progenitor from the pre-explosion photometry.  The agreement between the mass derived from the life-time of the young stellar population component around 2008bk and the mass derived by \citet{2014MNRAS.438.1577M} supports the higher mass interpretation for that SN progenitor, irrespective of the different values of extinction found.  For SNe~2012aw and 2013ej, where there is ambiguity in the possible mass of the progenitor from pre-explosion images, the stellar populations analysis consistently supports the high mass interpretation. Coupled with the higher levels of extinction inferred towards the populations, this may suggest that progenitor mass constraints from pre-explosion images may have systematically underestimated \citep{2015PASA...32...16S}.  In addition, while our analysis is not sensitive to the roles of dust occurring in the circumstellar environments of the progenitors, we can see for SNe such as SN~2013ej that other RSGs associated with the progenitor's age component show a large scatter in brightness that is not reflected in the rest of the population, especially on the blue plume.  The study by \citet{2016MNRAS.463.1269B} showed that the degree of reddening arising in the CSM increases towards the end of the RSGs phase, and can lead to significant underestimates of the progenitor's luminosity (and, hence, initial mass).  This additional extinction could also explain those progenitors in our analysis which appear to lie between stellar populations.   In the case of SN~2003gd, the extinction derived towards the population is only $A_{V} = 0.4\,\mathrm{mags}$, however the analysis of the pre-explosion SED showed that the progenitor could be subject to as much as $A_{V} \sim 2.5\,\mathrm{mags}$, with a maximum initial mass of $\sim 15M_{\odot}$ consistent with the age of the young stellar population component.

The Bayesian mixture model we employ does not necessarily provide a single age solution for an individual star and the weight values $w$ in Table \ref{tab:res} correspond to the stellar demographics and can be converted, using the Monte Carlo simulations presented in Figure \ref{fig:disc:compplot}, into a total stellar mass for each age component. Caution is, therefore, required in interpreting these values in the context of a single CMD, which corresponds to only a single slice through the complex observed photometric data space.  In ideal circumstances, one would have multi-band photometry covering UV to near-infrared wavelengths to make a complete parameterisation of the properties of each star and, possibly, consider the population on the Hertzsprung-Russell diagram instead; however this would still be limited for those objects at the hottest and coolest temperature limits, which would still be subject to poor photometric coverage at the bluest and reddest extremes of the data space.   Even considering a wide wavelength range, the reliance on broad-band photometry is insensitive to the emission line features that may be associated with evolved massive stars such as WR stars, for example, for which broad-band photometry is contaminated by strong emission lines \citep{2007ARA&A..45..177C} but the Padova isochrones approximate the SEDs by simple black body functions \citep{2002A&A...391..195G}.

For our analysis we have assumed a \citet{1955ApJ...121..161S} IMF with $\alpha = 2.35$.  In circumstances where there is ambiguity as to which mass point might best describe a star (i.e. if the isochrone loops over itself in magnitude space), the prior on the initial mass of the star (i.e. the IMF) weights the analysis in favour of the lower mass solution.   We note that other forms of the IMF \citep[e.g.][]{2001ApJ...554.1274C,2001MNRAS.322..231K,2002Sci...295...82K} have similar values for $\alpha$ in the high mass regime ($M > 1M_{\odot}$) which our observations are sensitive to.  As noted by \citet{2002Sci...295...82K}, these forms of the IMF may be biassed for high mass stars by unresolved binaries, which may imply that the true IMF is in fact steeper.  To assess the dependency of our analysis on the choice of IMF and uncertainties on the value of $\alpha$, we consider the $1\sigma$ errors bars on $\alpha$ reported by \citet{2001MNRAS.322..231K} and reanalysed our observations of the population around SN~2008bk for $\alpha = 1.6$ and $3.0$.  There is significant support for the steeper IMF ($\alpha = 3.0$) with $\ln K (\mathrm{Kroupa\,\alpha=3.0, Salpeter}) = 51.3$, but this results in only a marginal change in the derived parameters over the Salpeter IMF with the age of the youngest population changing by $\Delta \tau = +0.03$ and the extinction decreasing by $\Delta A_{V} = -0.02\,\mathrm{mags}$.  The shallower Kroupa IMF is less preferred than the Salpeter IMF with $\ln K (\mathrm{Kroupa\,\alpha=1.6, Salpeter}) = -123.7$, and results in slightly younger redder solutions, although again within the uncertainties established with the Salpeter IMF.

A key outstanding problem associated with the youngest isochrones and the brightest blue stars is possible contamination by clusters.  One precaution against contamination is to exclude those objects that {\sc dolphot} determines to be extended, however the sharpness and $\chi^{2}$ parameters may not be sufficient to appropriately diagnose the spatial extent of sources.  An alternative strategy might  be to employ packages such as {\sc ishape} \citep{1999A&AS..139..393L} and {\sc galfit} \citep{2010AJ....139.2097P}, that are specifically designed to explore the spatial extent of sources.  Unfortunately, the ability to identify barely resolved compact clusters diminishes with distance, which may particularly affect our analysis of the population around SN~2009kr, such that additional constraints from brightness and colour will also be required to classify the brightest sources in the CMD.  The suggestion of \citet{2005A&A...443...79B} that sources with $M_{V} < -8.6\,\mathrm{mags}$ are likely to be clusters cannot be used to exclude bright sources from this analysis, since this would particularly impact massive stars on the youngest isochrones which are brighter than this constraint.  The problem is further compounded at the lower cluster mass range, where the final brightness and colours of clusters may have a wide range of values for a given mass, due to the dependence on the cluster SED on a small number of bright stars.   A future extension of the work presented here would consider possible contamination by clusters into the Bayesian framework used here, but it requires a proper generative model for the sizes and shapes of clusters (including the instrumental PSF) to interpret parameters such as the concentration index $C$ \citep[e.g.][]{1999AJ....118.1551W, 2010ApJ...719..966C,  2016ApJ...824...71C}.  In the absence of useful spatial constraints on the possible cluster-like nature of sources, the key unknown is the expected relative prior probability that a given source might be a cluster ($p\left(\mathrm {cluster}\right)  = 1 - p\left(\mathrm{star}\right) $) and how it depends on the spatial scales over which massive stars form.

%%%%%%%%%%%%%%%%%%%%%%%%%%%%%%%%
%CONCLUSION
%%%%%%%%%%%%%%%%%%%%%%%%%%%%%%%%
\section{Conclusions}
\label{sec:conc}
We have presented age constraints for the stellar populations  within a projected radius of $\mathrm{100\,\mathrm{pc}}$ of the sites of a sample of 12 extragalactic Type IIP SNe.  For all SN locations, we find that the local stellar population requires more than one age component to accurately fit the observed CMD.  We find, in the majority of cases, age components corresponding to the lifetimes of stars within $\sim \pm3M_{\odot}$ of the initial masses derived from direct detection of the progenitor in pre-explosion images.  The progenitor of SN~2004et, for which the interpretation of the pre-explosion observations is ambiguous, may match younger isochrones, implying a significantly higher mass for the progenitor than previously assumed.
We have found that the extinctions derived from the analysis of the stellar populations is $\sim 0.3\,\mathrm{mags}$ greater than the values previously applied to the direct analysis of the progenitors; but also find clear evidence for differential extinction giving rise to a range of extinction values, which may particularly affect RSGs.
The comparison between the analysis of the stellar population associated with and the pre-explosion observations of the progenitor of SN~2008bk yield approximately the same progenitor mass, although vastly different extinctions.   In the cases of four SNe in our sample, there exist in the same field stars that are younger and more massive than the star that exploded.

%%%%%%%%%%%%%%%%%%%%%%%%%%%%%%%%
%ACKNOWLEDGMENTS
%%%%%%%%%%%%%%%%%%%%%%%%%%%%%%%%
\section*{Acknowledgments}
The research of JRM is supported through a Royal Society University Research Fellowship.  Based on observations made with the NASA/ESA Hubble Space Telescope, obtained from the Data Archive at the Space Telescope Science Institute, which is operated by the Association of Universities for Research in Astronomy, Inc., under NASA contract NAS 5-26555. These observations are associated with programs 9796, 10607, 11675, 12574, 12262, 12285, 13364, 13392, 13825 and 14226. 
\bibliographystyle{mnras}

\end{document}